\documentclass[twocolumn,showpacs,preprintnumbers,amsmath,amssymb]{revtex4}
\usepackage{amsmath,amssymb,graphicx}
\usepackage{graphicx}
\usepackage{dcolumn}
\usepackage{bm}
\usepackage{epsfig}
\usepackage{here}
\usepackage{float}
\usepackage{amsmath}
\usepackage[usenames]{color}
\usepackage{amsmath}
\usepackage{epsfig}
\usepackage{color}

\newcommand{\bea}{\begin{eqnarray}}
\newcommand{\eea}{\end{eqnarray}}

\input{tcilatex}

\begin{document}


\title{Perturbations and Linearization Stability of Closed Friedmann Universes}
\author{Hyerim Noh${}^{1,2}$, Jai-chan Hwang${}^{3,2}$ and John D. Barrow$%
{}^{2}$}
\address{${}^{1}$Center for Large Telescope, Korea Astronomy and Space
         Science Institute, Daejon, Republic of Korea \\
         ${}^{2}$Centre for Theoretical Cosmology, DAMTP,
         University of Cambridge, CB3 0WA Cambridge, United Kingdom \\
         ${}^{3}$Department of Astronomy and Atmospheric Sciences,
         Kyungpook National University, Daegu, Republic of Korea}

\date{\today}

\begin{abstract}
We consider perturbations of closed Friedmann universes.
Perturbation modes of two
lowest wavenumbers ($L=0$ and $1$) are generally known to be fictitious, but
here we show that both are physical. The issue is more subtle in Einstein
static universes where closed background space has a time-like Killing
vector with the consequent occurrence of linearization instability.
Proper solutions of the linearized equation need to satisfy the Taub constraint on
a quadratic combination of first-order variables. We evaluate the Taub
constraint in the two available fundamental gauge conditions, and show that
in both gauges the $L\geq 1$ modes should accompany the $L=0$ (homogeneous)
mode for vanishing sound speed, $c_{s}$. For $c_{s}^{2}>1/5$
(a scalar field supported Einstein static model belongs to this case
with $c_s^2 = 1$), the $L\geq 2$
modes are known to be stable. In order to have a stable Einstein static
evolutionary stage in the early universe, before inflation and without
singularity, although the Taub constraint does not forbid it, we need to
find a mechanism to suppress the unstable $L=0$ and $L=1$ modes.
\end{abstract}

\pacs{04.20.-q, 04.25.Nx, 98.80.-k, 98.80.Jk}
\maketitle


%
%

\section{Introduction}

The study of scalar, vector and tensor perturbations of the Friedmann
universes of general relativity began with the famous paper of Lifshitz in
1946 \cite{Lifshitz-1946, Lifshitz-Khalatnikov-1963}. Curiously, just as the
isotropic and homogeneous Newtonian cosmologies were found later by Milne
and McCrea \cite{milne}, in 1934, than their general relativistic
counterparts by Friedmann \cite{Friedmann-1922, Friedmann-1924} in 1922,
so the Newtonian treatment of their scalar perturbations, by
Bonnor \cite{bonn} in 1957, also followed the general relativistic treatment
of Lifshitz. Recently, the studies of the stability of the Einstein static
universe by Barrow et al \cite{Barrow-etal-2003} and Losic and Unruh \cite%
{Losic-Unruh-2005} have drawn attention to a subtle feature of the
homogeneous and isotropic background cosmological model that can cause
perturbation theory to fail due to the phenomenon of linearisation
instability. This is the motivation for our study.

Linearisation instability arises when the sum of the two leading terms in
perturbation around an exact solution cannot be completed to a convergent
expansion. That is, if the metric is expanded as
\begin{equation}
g_{ab}=g_{ab}^{(0)}+\epsilon g_{ab}^{(1)}+\epsilon ^{2}g_{ab}^{(2)}+\dots ,
\label{a}
\end{equation}%
where $g_{ab}^{{}}$ and $g_{ab}^{(0)}$ are solutions of the full Einstein
equations, and $g_{ab}^{(1)}$ is a solution of the linearised Einstein
equations, then the series expansion is said to be \textit{linearisation
stable} if the series (\ref{a}) can be completed to form a convergent
series. If not, it is said to be \textit{linearisation unstable}. In general
relativity, Fischer and Marsden, Arms, and Moncrief \cite{FMM, mon, arms}
showed that compact spaces in vacuum with Killing vectors are linearisation
unstable: $g_{ab}^{(1)}$
is linearisation stable if and only if $g_{ab}^{(0)}$ has no Killing fields.
In ref. \cite{BT}, this feature was discussed in relation to series
expansions about the Mixmaster universe, which has compact space sections
and Killing symmetries, and Brill provides several examples \cite{brill}. A
comprehensive overview \emph{is also given in the thesis of }Altas \cite%
{altas}.

Heuristically, the geometry of the solution space of cosmologies with
compact Cauchy surfaces is conical at the points with Killing symmetries and
so the perturbation expansion is like trying to draw a tangent through the
apex of a cone: there are an infinite number of possible tangents and the
ones that form the leading order of an expansion that converges to a true
solution corresponds to the tangents that run down the side of the cone.
This reminds us that there are two ways to obtain a perturbed version of an
exact solution. The first (definitive but unrealistic) method is to find the
general solution\emph{\ of the equations} and linearise about the exact
solution in question. The other method (used in practice) is to linearise
the equations about the exact solution and\emph{\ solve the linearised
equations}. This does not necessarily lead to the same result unless some
extra constraints are \emph{imposed.} (which we shall discuss below in the
general relativistic context).

A typical example is provided by the equation,
\begin{equation}
f(x,y)=x(x^{2}+y^{2})= 0,  \label{b}
\end{equation}%
%
with the set of solutions $(x,y)=(0,y)$, \emph{where }$y$\emph{\ is arbitrary%
}. Now linearise Eq.\ (\ref{b}) about the particular solution $(0,0)$.
This yields
\begin{equation}
(3x^{2}+y^{2})\delta x+2xy\delta y=0.  \label{c}
\end{equation}%
We see that for $(x,y)=(0,0)$ there is no restriction on $\ $the linearised
solutions and $(\delta x,\delta y)$ are completely arbitrary. However, from
the exact solution, we know that although there are linearised solutions to
the linearised Eq.\ (\ref{c}) with $\delta x\neq 0$, they cannot arise
from \emph{the linearisations of any exact solution of} Eq.\ (\ref{b}) \cite%
{altas}.

Fischer, Marsden and Moncrief \cite{FMM} showed that the $g_{ab}^{(1)}$ is
not a spurious solution if and only if it satisfies a second-order
constraint, involving integrals of the Taub conserved quantity which
therefore vanish \cite{mon1974}. In this paper, we will evaluate the Taub
constraint in different gauges and determine the status of the first-order
neutral stability results for the Einstein static universe, \emph{which is a
prime candidate for the phenomenon of linearisation instability as it has
compact space sections and many Killing symmetries. }

In the course of this analysis we will also identify some features of gauge
invariant perturbation claims \emph{in the literature} that appear to be
discrepant in ways that do appear to have been noticed in the past.
Specifically, we will address two issues in the cosmological scalar
perturbations of the homogeneous and isotropic Friedmann universes. In \emph{%
some of} the literature, the perturbations with the two lowest wave numbers (%
$L=0$ and $1$) are claimed to be fictitious. Here, we show that both are
physical.

In a closed background space with Killing vectors,\emph{\ in order to be
linearisation stable} the solution of linearized equation should satisfy a
constraint on a quadratic combination of first-order variables, we call it
the Taub constraint. When the Taub constraint is evaluated under two gauge
conditions for a timelike Killing vector in an Einstein static background,
it implies that $L\geq 1$ modes should accompany the $L=0$ (homogeneous)
mode, but this is true only for vanishing sound speed.

In Section \ref{Linear-perturbation}, we review the equations and solutions
for linear perturbations of scalar type in the presence of background
curvature and we consider a complete set of exact solutions with
zero-pressure and cosmological constant (see the Appendix \ref%
{App:MDE-solutions}).

In Section \ref{Physical-nature} we investigate the physical nature of the
two lowest wave number modes ($L=0$ and $1$) in the positive curvature
background. In Section \ref{Einstein-static} we analyze the stability in the
Einstein static background in the presence of pressure or a scalar field. In
Section \ref{Taub-constraint} we evaluate the Taub constraint for a timelike
Killing vector in the Einstein static model; the Taub constraint is derived
in the Appendix \ref{App:Taub-constraint}. Section \ref{Discussion} is a
discussion of our results and their consequences. In Sections \ref%
{Linear-perturbation}-\ref{Einstein-static}, we consider scalar-type linear
perturbation in the Friedmann background with spatial curvature, while
Section \ref{Taub-constraint} considers second-order perturbations. Sections %
\ref{Linear-perturbation} and \ref{Physical-nature} consider general
background curvature $K$, while Sections \ref{Einstein-static} and \ref%
{Taub-constraint} are concerned with the positive curvature background. In
the case of the scalar field we set $c\equiv 1\equiv \hbar $.

%
%

\section{Linear perturbations with general curvature}

\label{Linear-perturbation}

All results in this section are known in the literature, but we pay special
attention to three simple cases of perturbed Friedmann universes. These are
(i) the Einstein static background with $H=0$, (ii) the homogeneous
perturbation with $\Delta =0$, and (iii) the case with $\Delta +3K=0$, where
$K$ is the curvature parameter in the Friedmann equation, equal to $0$ or $%
\pm 1$; the latter two cases are considered in the spherical geometry;
$H$ is the Hubble parameter and $\Delta$ is
a Laplacian operator of the comoving three-space of the Friedmann metric.
In these simple cases some terms in the perturbation equations automatically
vanish, thus the analysis and final results are often invalid; in such cases
a simple cure is to go back to the original perturbation equations and check
each case. We will study these simple cases in more detail in later Sections.

\begin{widetext}


\subsection{Basic equations}

We consider perturbations of scalar-type in the Friedmann background. Our
metric convention follows Bardeen's in \cite{Bardeen-1988}
\begin{equation}
\widetilde{g}_{00}=-a^{2}\left( 1+2\alpha \right) ,\quad \widetilde{g}%
_{0i}=-a^{2}\beta _{,i},\quad \widetilde{g}_{ij}\equiv a^{2}\left[ \left(
1+2\varphi \right) \gamma _{ij}+2\gamma _{,i|j}\right] ,  \label{metric}
\end{equation}%
with $x^{0}=\eta $ the conformal time and $a(t)$ is the expansion scale
factor. We introduce $\chi \equiv a(\beta +{\frac{a}{c}}\dot{\gamma})$ where
the time derivative is with respect to $t$ where $cdt\equiv ad\eta $. The
energy-momentum tensor is decomposed into fluid quantities based on a
timelike fluid four-vector, $\widetilde{u}_{a}$, normalized with $\widetilde{%
u}^{a}\widetilde{u}_{a}\equiv -1$, so
\begin{eqnarray}
&&\widetilde{T}_{ab}=\widetilde{\mu }\widetilde{u}_{a}\widetilde{u}_{b}+%
\widetilde{p}\left( \widetilde{g}_{ab}+\widetilde{u}_{a}\widetilde{u}%
_{b}\right) +\widetilde{\pi }_{ab}, \\
&&\widetilde{\mu }=\widetilde{\varrho }c^{2},\quad \widetilde{\varrho }%
\equiv \varrho +\delta \varrho ,\quad \widetilde{p}=p+\delta p,\quad
\widetilde{u}_{i}\equiv -{\frac{a}{c}}v_{,i},\quad \widetilde{\pi }%
_{ij}\equiv {\frac{1}{a^{2}}}\left( \nabla _{i}\nabla _{j}-{\frac{1}{3}}%
\gamma _{ij}\Delta \right) \Pi .
\end{eqnarray}%
We may set $\delta p\equiv c_{s}^{2}c^{2}\delta \varrho +e$ with $%
c_{s}^{2}\equiv \dot{p}/(\dot{\varrho}c^{2})$, and $w\equiv p/(\varrho c^{2})
$; $e$ is the entropic perturbation and $\Pi $ is the anisotropic stress.
The spatial indices are raised and lowered by $\gamma _{ij}$ and its
inverse, and a vertical bar indicates the covariant derivative based on the
metric tensor $\gamma _{ij}$. One representation of $\gamma _{ij}$ is
\begin{equation}
d\ell ^{2}\equiv \gamma _{ij}dx^{i}dx^{j}=\left\{
\begin{array}{ll}
d\chi ^{2}+\sin ^{2}{\chi }\left( d\theta ^{2}+\sin ^{2}{\theta }d\phi
^{2}\right) ,\quad  & (K=+1) \\
d\chi ^{2}+\chi ^{2}\left( d\theta ^{2}+\sin ^{2}{\theta }d\phi ^{2}\right) ,
& (K=0) \\
d\chi ^{2}+\sinh ^{2}{\chi }\left( d\theta ^{2}+\sin ^{2}{\theta }d\phi
^{2}\right) , & (K=-1)%
\end{array}%
\right.   \label{spatial-metric}
\end{equation}%
with a normalized background curvature $K$.

The Friedmann equations are \cite{Friedmann-1922, Friedmann-1924}
\begin{eqnarray}
&&H^{2}={\frac{8\pi G}{3}}\varrho -{\frac{Kc^{2}}{a^{2}}}+{\frac{\Lambda
c^{2}}{3}},\quad \dot{H}+H^{2}=-{\frac{4\pi G}{3}}\left( \varrho +{\frac{3p}{%
c^{2}}}\right) +{\frac{\Lambda c^{2}}{3}},\quad \dot{H}=-4\pi G\left(
\varrho +{\frac{p}{c^{2}}}\right) +{\frac{Kc^{2}}{a^{2}}},  \notag \\
&&\dot{\varrho}+3H\left( \varrho +{\frac{p}{c^{2}}}\right) =0,  \label{BG}
\end{eqnarray}%
with $H \equiv \dot a/a$ the Hubble-Lema\^itre parameter.
The pressure term was first considered by Lema\^{\i}tre \cite{Lemaitre-1927,
Lemaitre-1931a,Lemaitre-1931b}. The general perturbation in the Friedmann
background was studied first by Lifshitz \cite{Lifshitz-1946,
Lifshitz-Khalatnikov-1963}. Lifshitz studied the scalar, vector and
tensor-type perturbations to the linear order in the synchronous gauge ($%
\alpha \equiv 0\equiv \beta $). Here we consider the scalar-type
perturbation in all fundamental gauges. Unless mentioned otherwise, our
study in Sections \ref{Linear-perturbation} and \ref{Physical-nature} is
valid for all type of $K$, whereas Sections \ref{Einstein-static} and \ref%
{Taub-constraint} concern the positive curvature model.

To linear order in perturbation, the basic equations for the scalar-type
perturbation, without imposing the gauge conditions, are \cite{Bardeen-1988}
\begin{eqnarray}
&&\kappa =3H\alpha -3\dot{\varphi}-c{\frac{\Delta }{a^{2}}}\chi ,
\label{eq1} \\
&&4\pi G\delta \varrho +H\kappa +c^{2}{\frac{\Delta +3K}{a^{2}}}\varphi =0,
\label{eq2} \\
&&\kappa +c{\frac{\Delta +3K}{a^{2}}}\chi -{\frac{12\pi G}{c^{2}}}a\left(
\varrho +{\frac{p}{c^{2}}}\right) v=0,  \label{eq3} \\
&&\dot{\kappa}+2H\kappa +\left( 3\dot{H}+c^{2}{\frac{\Delta }{a^{2}}}\right)
\alpha =4\pi G\left( \delta \varrho +{\frac{3\delta p}{c^{2}}}\right) ,
\label{eq4} \\
&&\varphi +\alpha -{\frac{1}{c}}\left( \dot{\chi}+H\chi \right) =-{\frac{%
8\pi G}{c^{4}}}\Pi ,  \label{eq5} \\
&&\delta \dot{\varrho}+3H\left( \delta \varrho +{\frac{\delta p}{c^{2}}}%
\right) +\left( \varrho +{\frac{p}{c^{2}}}\right) \left( 3H\alpha -\kappa -{%
\frac{\Delta }{a}}v\right) =0,  \label{eq6} \\
&&{\frac{1}{a^{4}}}\left[ a^{4}\left( \varrho +{\frac{p}{c^{2}}}\right) v%
\right] ^{\displaystyle\cdot }={\frac{1}{a}}\left[ \delta p+\left( \varrho
c^{2}+p\right) \alpha +{\frac{2}{3}}{\frac{\Delta +3K}{a^{2}}}\Pi \right] .
\label{eq7}
\end{eqnarray}

We consider a gauge transformation, $\widehat{x}^{c}=x^{c}+\widetilde{\xi }%
^{c}(x^{e})$ with $\widetilde{\xi }^{0}=\xi ^{0}\equiv {\frac{1}{a}}\xi ^{t}$
and $\widetilde{\xi }^{i}=\xi ^{i}\equiv {\frac{1}{a}}\xi ^{|i}$; index of $%
\xi _{i}$ is raised and lowered using $\gamma _{ij}$ as the metric. To the
linear order we have \cite{Bardeen-1988}
\begin{eqnarray}
&&\widehat{\alpha }=\alpha -{\frac{1}{c}}\dot{\xi}^{t},\quad \widehat{\beta }%
=\beta -{\frac{1}{a}}\xi ^{t}+{\frac{a}{c}}\left( {\frac{1}{a}}\xi \right) ^{%
\displaystyle{\cdot }},\quad \widehat{\gamma }=\gamma -{\frac{1}{a}}\xi
,\quad \widehat{\chi }=\chi -\xi ^{t},\quad \widehat{\kappa }=\kappa +{\frac{%
1}{c}}\left( 3\dot{H}+c^{2}{\frac{\Delta }{a^{2}}}\right) \xi ^{t},  \notag
\\
&&\widehat{\varphi }=\varphi -{\frac{1}{c}}H\xi ^{t},\quad \delta \widehat{%
\varrho }=\delta \varrho -{\frac{1}{c}}\dot{\varrho}\xi ^{t},\quad \delta
\widehat{p}=\delta p-{\frac{1}{c}}\dot{p}\xi ^{t},\quad \widehat{v}=v-{\frac{%
c}{a}}\xi ^{t},\quad \widehat{e}=e,\quad \widehat{\Pi }=\Pi ,\quad \delta
\widehat{\phi }=\delta \phi -{\frac{1}{c}}\dot{\phi}\xi ^{t}.  \label{GT}
\end{eqnarray}%
By using $\chi $ instead of $\beta $ and $\gamma $, all the perturbation
variables are spatially gauge invariant. We have the following possible
fundamental gauge conditions: the uniform-curvature gauge (UCG, $\varphi
\equiv 0$), the uniform-density gauge (UDG, $\delta \varrho \equiv 0$), the
uniform-expansion gauge (UEG, $\kappa \equiv 0$), the comoving gauge (CG, $%
v\equiv 0$), the zero-shear gauge (ZSG, $\chi \equiv 0$), and the
synchronous gauge (SG, $\alpha \equiv 0$). We introduce gauge-invariant
notations, like $v_{\chi }\equiv v-(c/a)\chi \equiv -(c/a)\chi _{v}$;  where
$v_{\chi }$ is gauge invariant with the \textit{same} as $v$ in the ZSG. One
exception is the SG; after imposing the gauge condition we still have
non-vanishing $\xi ^{t}(\mathbf{x})$ which is the remnant gauge mode in the
SG. Thus, $\chi _{\alpha }\equiv \chi -c\int^{t}\alpha dt$ is not gauge
invariant; the lower bound of integration gives the remnant gauge mode with $%
\chi \propto \xi ^{t}(\mathbf{x})$. Concerning the spatial gauge
transformation, our definitions of $\chi $ and $v$ are spatially
gauge-invariant combinations; $\chi $ is the same as (equivalent to) $a\beta
$ under the spatial gauge condition $\gamma \equiv 0$.

We note that in a static background with $H=0$, both $\varphi $ and $\delta
\varrho $ become gauge-invariant. In addition, for $\Delta =0$ (thus, a
homogeneous) mode, $\kappa $ becomes gauge invariant as well.


\subsection{Exact equations and asymptotic solutions}

\label{sec:Phi-eqs}

A powerful large-scale conserved behavior of a combination of variables in
the presence of $K$ is known already. The following analysis is valid for $%
H\neq 0$; for $H=0$, $\varphi $ and $\delta $ are gauge invariant, and we
can show $\Phi =0$, (see Section \ref{Einstein-static}). We define
\begin{equation}
\Phi \equiv \varphi _{v}-{\frac{{\frac{Kc^{2}}{a^{2}}}}{4\pi G\left( \varrho
+{\frac{p}{c^{2}}}\right) }}\varphi _{\chi }=\varphi _{v}+{\frac{K}{\Delta
+3K}}{\frac{\delta _{v}}{1+w}},  \label{Phi-definition}
\end{equation}%
where we used
\begin{equation}
c^{2}{\frac{\Delta +3K}{a^{2}}}\varphi _{\chi }=-4\pi G\delta \varrho _{v},
\label{Poisson-eq}
\end{equation}%
which follows from Eqs.\ (\ref{eq2}) and (\ref{eq3}); from Eq.\ (\ref{GT})
we have
\begin{equation*}
\varphi _{\chi }\equiv \varphi -{\frac{H}{c}}\chi \text{ \ and \ }\delta
\varrho _{v}\equiv \delta \varrho +{\frac{1}{c^{2}}}3aH\left(\varrho +\frac{p%
}{c^{2}}\right)v.
\end{equation*}
Note that for $\Delta =-3K$, Eq.\ (\ref{Poisson-eq}) gives $\delta \varrho
_{v}=0$, and the second expression in Eq.\ (\ref{Phi-definition}) does not
apply; $\delta \varrho _{v}=0$ follows from Eqs.\ (\ref{eq2}) and (\ref{eq3}%
) evaluated in the CG with $\Delta =-3K$.

In Eq.\ (\ref{Phi-definition}), from the first relation, using Eqs.\ (\ref%
{eq1}), (\ref{eq3}) and (\ref{eq5}), and from the second relation, using
Eqs.\ (\ref{eq1})-(\ref{eq3}), (\ref{eq6}) and (\ref{eq7}), respectively, we
can derive
\begin{eqnarray}
&&\Phi ={\frac{H^{2}}{4\pi G\left( \varrho +{\frac{p}{c^{2}}}\right) a}}%
\left[ \left( {\frac{a}{H}}\varphi _{\chi }\right) ^{\displaystyle{\cdot }}+{%
\frac{8\pi G}{c^{4}}}a\Pi \right] ,  \label{Phi-eq1} \\
&&\dot{\Phi}={\frac{Hc_{s}^{2}c^{2}}{4\pi G\left( \varrho +{\frac{p}{c^{2}}}%
\right) }}{\frac{\Delta }{a^{2}}}\varphi _{\chi }-{\frac{H}{\varrho c^{2}+p}}%
\left( e+{\frac{2}{3}}{\frac{\Delta }{a^{2}}}\Pi \right) .  \label{Phi-eq2}
\end{eqnarray}%
Although we used Eq.\ (\ref{Poisson-eq}) in deriving Eq.\ (\ref{Phi-eq2}),
we can check by using the original Eqs.\ (\ref{eq1})-(\ref{eq7}) that the
result is valid even for $\Delta =-3K$. Ignoring the imperfect fluid
contribution, thus setting $e\equiv 0\equiv \Pi $, we have
\begin{equation}
{\frac{H^{2}c_{s}^{2}}{\left( \varrho +{\frac{p}{c^{2}}}\right) a^{3}}}\left[
{\frac{\left( \varrho +{\frac{p}{c^{2}}}\right) a^{3}}{H^{2}c_{s}^{2}}}\dot{%
\Phi}\right] ^{\displaystyle{\cdot }}-c_{s}^{2}c^{2}{\frac{\Delta }{a^{2}}}%
\Phi =0.
\end{equation}%
Using
\begin{equation}
v\equiv z\Phi ,\quad z\equiv {\frac{a\sqrt{\varrho +{\frac{p}{c^{2}}}}}{%
Hc_{s}}},
\end{equation}%
we have
\begin{equation}
{\frac{1}{c^{2}az^{2}}}\left( az^{2}\dot{\Phi}\right) ^{\displaystyle{\cdot }%
}-c_{s}^{2}{\frac{\Delta }{a^{2}}}\Phi ={\frac{1}{a^{2}z}}\left[ v^{\prime
\prime }-\left( {\frac{z^{\prime \prime }}{z}}+c_{s}^{2}\Delta \right) v%
\right] =0,  \label{v-eq}
\end{equation}%
where a prime is the time derivative with respect to the conformal time, $%
\eta $. In the large-scale (super sound-horizon scale) limit, $z^{\prime
\prime }/z\gg c_{s}^{2}\Delta $, we have a general solution:
\begin{equation}
\Phi (\mathbf{x},t)=C(\mathbf{x})+d(\mathbf{x})\int^{t}{\frac{H^{2}c_{s}^{2}%
}{4\pi G\left( \varrho +{\frac{p}{c^{2}}}\right) a^{3}}}dt.
\label{Phi-solution}
\end{equation}%
Thus, the relatively growing solution of $\Phi $ remains constant in the
super-sound-horizon scale. The sound-horizon vanishes for zero-pressure
fluid, in which case we have $\dot{\Phi}=0$, and so $\Phi =C(\mathbf{x})$
exactly.

The well known equation in terms of $v$ and $z$ in Eq.\ (\ref{v-eq}) first
appeared in Eq.\ (44) of Field and Shepley's 1968 paper \cite%
{Field-Shepley-1968} in the context with general $K$ (see also \cite%
{Chibisov-Mukhanov-1982}, Section V of \cite{Hwang-Vishniac-1990} and
Section III of \cite{Hwang-Noh-2005}; in the absence of $K$, see \cite%
{Lukash-1980a, Lukash-1980b, Mukhanov-1988}). Using Eqs.\ (\ref{eq2}) and (%
\ref{eq6}), Eq.\ (\ref{Phi-definition}) can be arranged as
\begin{equation}
c^{2}{\frac{\Delta +3K}{a^{2}H^{2}}}\Phi =-\left( {\frac{\delta \varrho
_{\alpha }}{\left( \varrho +{\frac{p}{c^{2}}}\right) H}}\right) ^{%
\displaystyle{\cdot }}-{\frac{3e}{\varrho c^{2}+p}},  \label{Phi-SG}
\end{equation}%
which is related to Eqs.\ (31) and (43) in \cite{Field-Shepley-1968}. Here,
we used $\alpha _{v}\equiv \alpha -{\frac{1}{c^{2}}}\left( av\right) ^{%
\displaystyle{\cdot }}$, $v_{\alpha }\equiv v-{\frac{c^{2}}{a}}%
\int^{t}\alpha dt$ and $\delta \varrho _{\alpha }\equiv \delta \varrho -\dot{%
\varrho}\int^{t}\alpha dt$ which follow from Eq.\ (\ref{GT}); notice that
the remnant gauge degree of freedom in the SG imbedded in the lower bound
of integration of $\delta \varrho _{\alpha }$ in Eq.\ (\ref{Phi-SG})
disappears because of the time derivative. For $\Delta =-3K$ Eq.\ (%
\ref{Phi-SG}) is identically satisfied as we have $\delta \varrho _{\alpha
}=\delta \varrho _{v}-\dot{\varrho}\int^{t}\alpha _{v}dt$ with $\delta
\varrho _{v}=0$ and $\alpha _{v}=-e/(\varrho c^{2}+p)$ which follows from
Eq.\ (\ref{eq7}).


\subsection{Exact solutions for zero-pressure fluid}

\label{MDE-solutions}

In the zero-pressure situation, with $p=0=\delta p$ and $\Pi =0$, but with
general $K$ and $\Lambda $, we have [see Eqs.\ (\ref{Phi-eq2}) and (\ref%
{Phi-solution})]:
\begin{equation}
\Phi =C(\mathbf{x}).
\end{equation}%
Again, the following analysis is valid for $H\neq 0$; for $H=0$, $\varphi $
and $\delta $ are gauge invariant, and we have $\Phi =0$; the static case
will be studied in Section \ref{Einstein-static}. In the CG, Eq.\ (\ref{eq7}%
) gives $\alpha _{v}=0$. Using Eqs.\ (\ref{eq2}) and (\ref{eq6}), the second
relation in Eq.\ (\ref{Phi-definition}) gives
\begin{equation}
\left( {\frac{\delta _{v}}{H}}\right) ^{\displaystyle{\cdot }}=-c^{2}{\frac{%
\Delta +3K}{a^{2}H^{2}}}\Phi ,
\end{equation}%
with an exact solution:
\begin{equation}
\delta _{v}=-c^{2}(\Delta +3K)CH\int^{t}{\frac{dt}{\dot{a}^{2}}}.
\label{delta_v-solution}
\end{equation}%
The relatively decaying solution is absorbed in the lower bound of the
integration. From this one solution we can derive all the other solutions in
the same gauge and, using the complete solutions in one gauge, we can derive
all solutions with all other gauge conditions. The complete solutions are
presented in Table 1 of \cite{Hwang-1994}, and are reproduced in the
Appendix \ref{App:MDE-solutions} to this paper, in our notation, paying
particular attention to the $k^{2}=0$ and $3K$ modes in the spherical
geometry; we introduced the comoving wave number with $\Delta =-k^{2}$.

For $\Delta =-3K$, we have $\delta _{v}=0$ and we cannot begin with above
two equations which become trivial. We need to start from a non-vanishing
solution. In the ZSG, from Eqs.\ (\ref{eq1}), (\ref{eq3}), (\ref{eq5}) and (%
\ref{eq7}), we have
\begin{equation}
{\frac{1}{a}}\left( a\varphi _{\chi }\right) ^{\displaystyle{\cdot }}=-{%
\frac{4\pi G\varrho }{c^{2}}}av_{\chi },\quad {\frac{1}{a}}\left( av_{\chi
}\right) ^{\displaystyle{\cdot }}=-{\frac{c^{2}}{a}}\varphi _{\chi },\quad
\mathrm{thus}\quad {\frac{1}{a^{3}}}\left[ a^{2}\left( a\varphi _{\chi
}\right) ^{\displaystyle{\cdot }}\right] ^{\displaystyle{\cdot }}=4\pi
G\varrho \varphi _{\chi }.
\end{equation}%
This can be written as
\begin{equation}
{\frac{1}{a^{3}H}}\left[ a^{2}H^{2}\left( {\frac{a}{H}}\varphi _{\chi
}\right) ^{\displaystyle{\cdot }}\right] ^{\displaystyle{\cdot }}=0,
\end{equation}%
with the solution
\begin{equation}
\varphi _{\chi }=4\pi G\varrho a^{2}HC\int^{t}{\frac{dt}{\dot{a}^{2}}}.
\label{varphi_chi-sol}
\end{equation}%
The normalization is made using Eq.\ (\ref{Phi-eq1}). This solution
coincides with the one in the Appendix \ref{App:MDE-solutions}. From this we
obtain solutions of every variable in all gauge conditions. The results are
naturally (because $\varphi _{\chi }$ coincides) the same as the ones
derived from $\delta _{v}$ in Eq.\ (\ref{delta_v-solution}) presented in the
Appendix \ref{App:MDE-solutions}.


\subsection{Scalar fields}

For a minimally coupled scalar field, the equations for the fluid, Eq.\ (\ref%
{BG}) for the background, and equations (\ref{eq1})-(\ref{eq7}) for
perturbations, remain valid with the fluid quantities replaced by the ones
for the scalar field. Additionally, we have the scalar field equation of
motion which also follows from the conservation equations, the last one in
Eq.\ (\ref{BG}) and Eq.\ (\ref{eq6}). For the background, we have
\begin{eqnarray}
&&\varrho ={\frac{1}{2}}\dot{\phi}^{2}+V,\quad p={\frac{1}{2}}\dot{\phi}%
^{2}-V,  \label{BG-fluid-MSF} \\
&&\ddot{\phi}+3H\dot{\phi}+V_{,\phi }=0.  \label{BG-EOM}
\end{eqnarray}%
For the perturbation, we have
\begin{eqnarray}
&&\delta \varrho =\dot{\phi}\delta \dot{\phi}-\dot{\phi}^{2}\alpha +V_{,\phi
}\delta \phi ,\quad \delta p=\dot{\phi}\delta \dot{\phi}-\dot{\phi}%
^{2}\alpha -V_{,\phi }\delta \phi ,\quad \left( \varrho +p\right) v ={\frac{1%
}{a}}\dot{\phi}\delta \phi ,\quad \Pi =0,  \label{pert-fluid-MSF} \\
&&\delta \ddot{\phi}+3H\delta \dot{\phi}+\left( V_{,\phi \phi }-{\frac{%
\Delta }{a^{2}}}\right) \delta \phi =\dot{\phi}\left( \kappa +\dot{\alpha}%
\right) +\left( 2\ddot{\phi}+3H\dot{\phi}\right) \alpha .  \label{pert-EOM}
\end{eqnarray}%
The gauge transformation property of the scalar field is presented in Eq.\ (%
\ref{GT}).

The scalar field can be treated as a fluid, as identified in Eqs.\ (\ref%
{BG-fluid-MSF}) and (\ref{pert-fluid-MSF}). The CG ($v\equiv 0$) coincides
with the uniform-field gauge (UFG, $\delta \phi \equiv 0$). In this gauge we
have
\begin{equation}
\delta \phi _{v}=\delta \mu _{v},\quad \mathrm{thus}\quad e=\left(
1-c_{s}^{2}\right) \delta \mu _{v}\quad \mathrm{with}\quad c_{s}^{2}\equiv {%
\frac{\dot{p}}{\dot{\varrho}}}=-1-{\frac{2\ddot{\phi}}{3H\dot{\phi}}}.
\end{equation}%
Using this $e$ and Eq.\ (\ref{Poisson-eq}), Eq.\ (\ref{Phi-eq2}) gives
\begin{equation}
\dot{\Phi}={\frac{Hc_{A}^{2}c^{2}}{4\pi G\left( \varrho +{\frac{p}{c^{2}}}%
\right) }}{\frac{\Delta }{a^{2}}}\varphi _{\chi }\quad \mathrm{with}\quad
c_{A}^{2}\Delta \equiv \Delta +3(1-c_{s}^{2})K.  \label{Phi-eq2-MSF}
\end{equation}%
Thus, the equations in Section \ref{sec:Phi-eqs} with $e=0=\Pi $ are valid
with $c_{s}^{2}$ replaced by $c_{A}^{2}$, see Section III of \cite%
{Hwang-Noh-2005}; in the absence of $K$, we have $c_{A}^{2}=1$ \cite%
{Mukhanov-1988}.

%
%
%

\section{The physical nature of the $k^{2}=0$ and $k^{2}=3K$ modes}

\label{Physical-nature}

In the spherical geometry the mode function has discrete wave numbers with $%
k^{2}\equiv (n^{2}-1)K$ and $n=1,2,\dots $; we often keep $K$ explicitly
even though we normalized earlier it as $K=1$. In the literature the two
lowest wave numbers with $n=1$ and $2$, thus $k^{2}=0$ and $3K$, are claimed
to be fictitious perturbations \cite{Lifshitz-1946,
Lifshitz-Khalatnikov-1963, Bardeen-1980}. Our review in the previous section
shows no particular trouble for $\Delta =0$ and $-3K$ cases. In this section
we study the individual case in more detail and show the physical
non-fictitious nature of these two modes.\emph{\ }


\subsection{$k^{2}=0$ (homogeneous) modes}

\label{sec:k^2=0}

First, we consider the $k^{2}=0$ mode. By setting $\Delta =0$ our basic
equations in (\ref{eq1})-(\ref{eq7}) become a set of ordinary differential
equations depending only on time and are therefore spatially homogeneous.

For $\Delta =0$ we have $\Phi =\varphi +\delta /[3(1+w)]$ and from Eqs.\ (%
\ref{eq1}) and (\ref{eq6}) we can show
\begin{equation}
\left( \varphi +{\frac{\delta }{3(1+w)}}\right) ^{\displaystyle{\cdot }}+{%
\frac{He}{\varrho c^{2}+p}}=0.
\end{equation}%
Thus, for $e=0$, we have
\begin{equation}
\varphi +{\frac{\delta }{3(1+w)}}=C.
\end{equation}

In the UDG ($\delta \equiv 0$), which is possible for $H\neq 0$, we have
(for $e=0$)
\begin{equation}
\varphi _{\delta }\equiv \varphi +{\frac{\delta }{3(1+w)}}=C.
\end{equation}%
From Eqs.\ (\ref{eq6}) and (\ref{eq2}), we have
\begin{equation}
\kappa _{\delta }=-{\frac{3Kc^{2}}{a^{2}H}}C,\quad \alpha _{\delta }=-{\frac{%
Kc^{2}}{a^{2}H^{2}}}C,
\end{equation}%
and the solutions for other variables in the same gauge follow from Eqs.\ (%
\ref{eq5}) and (\ref{eq3}):
\begin{eqnarray}
&&\chi _{\delta }={\frac{c}{a}}\int^{t}a\left[ \left( 1-{\frac{Kc^{2}}{%
a^{2}H^{2}}}\right) C+{\frac{8\pi G}{c^{4}}}\Pi \right] dt,  \notag \\
&&{\frac{4\pi G}{c^{2}}}\left( \varrho +{\frac{p}{c^{2}}}\right) av_{\delta
}=-{\frac{Kc^{2}}{a^{2}H}}C+{\frac{Kc^{2}}{a^{3}}}\int^{t}a\left[ \left( 1-{%
\frac{Kc^{2}}{a^{2}H^{2}}}\right) C+{\frac{8\pi G}{c^{4}}}\Pi \right] dt.
\end{eqnarray}

As we have solutions for a complete set of variables in the UDG, the
solutions in any other gauge can be derived using the gauge transformation
properties in Eq.\ (\ref{GT}). As an example, density perturbation in the
CG, $\delta _{v}$, and the curvature perturbation in the ZSG, $\varphi
_{\chi }$, can be derived in the following way. From Eq.\ (\ref{GT}) we have
\begin{equation}
\delta _{v}\equiv \delta +{\frac{3}{c^{2}}}aH\left( 1+w\right) v\equiv {%
\frac{3}{c^{2}}}aH\left( 1+w\right) v_{\delta },\quad \varphi _{\chi }\equiv
\varphi -{\frac{1}{c}}H\chi =\varphi _{\delta }-{\frac{1}{c}}H\chi _{\delta
}.
\end{equation}%
For comparison with exact solutions in the zero-pressure case presented in
Section \ref{MDE-solutions} and the Appendix \ref{App:MDE-solutions}, it is
convenient to have
\begin{equation}
{\frac{a}{H}}-\int^{t}a\left( 1-{\frac{Kc^{2}}{\dot{a}^{2}}}\right) dt=4\pi
G\varrho a^{3}\int^{t}{\frac{dt}{\dot{a}^{2}}}.
\end{equation}

For a static background we have $H=0$ so we have to go back to the original
equations in (\ref{eq1})-(\ref{eq7}) and the gauge transformation properties
in Eq.\ (\ref{GT}); as $\delta $ is naturally gauge invariant, we cannot
take the UDG. Although we cannot construct $\varphi _{\delta }$, the
combination $\varphi +\delta /[3(1+w)]$ is fine and still gives $C$, which
in fact vanishes, see Eq.\ (\ref{eq2}). The static situation will be studied
in Section \ref{ES-k=0}.


\subsection{$k^{2}=3K$ modes}

For $k^{2}=3K$ we have $\Delta +3K=0$. As we have $\Delta +3K$ terms often
appearing in our basic equations, many variables vanish in some gauges.
Following Bardeen \cite{Bardeen-1980} we consider the case of the UEG with $%
\kappa \equiv 0$. Eqs.\ (\ref{eq2})-(\ref{eq4}) give $\delta =0$, $v=0$ and $%
\alpha =-e/(\varrho c^{2}+p)$, respectively; thus, the UDG and the CG also
give the identical results. Despite this simplification in the three gauges
(UEG, UDG and CG), Eqs.\ (\ref{eq1}) and (\ref{eq5}) give the following
equations:
\begin{eqnarray}
&&\dot{\varphi}={\frac{Kc}{a^{2}}}\chi -{\frac{He}{\varrho c^{2}+p}},\quad {%
\frac{1}{ca}}\left( a\chi \right) ^{\displaystyle{\cdot }}=\varphi -{\frac{e%
}{\varrho c^{2}+p}}+{\frac{8\pi G}{c^{4}}}\Pi ;  \notag \\
&&{\frac{1}{a^{3}}}\left[ a^{3}\left( \dot{\varphi}+{\frac{He}{\varrho
c^{2}+p}}\right) \right] ^{\displaystyle{\cdot }}-{\frac{Kc^{2}}{a^{2}}}%
\left( \varphi -{\frac{e}{\varrho c^{2}+p}}\right) ={\frac{8\pi G}{c^{2}}}{%
\frac{K}{a^{2}}}\Pi ,  \label{n=2-varphi-eq}
\end{eqnarray}%
which are valid for the three gauge conditions. For $e=0=\Pi $ we have
\begin{equation}
\dot{\varphi}={\frac{Kc}{a^{2}}}\chi ,\quad {\frac{1}{ca}}\left( a\chi
\right) ^{\displaystyle{\cdot }}=\varphi ;\quad {\frac{1}{a^{3}}}\left( a^{3}%
\dot{\varphi}\right) ^{\displaystyle{\cdot }}-{\frac{Kc^{2}}{a^{2}}}\varphi
=0.  \label{n=2-varphi-eq-2}
\end{equation}%
These are non-trivial equations and as the gauge modes are completely fixed
in all three gauges, the variables cannot be removed by gauge
transformation. Equations in the other remaining gauges (the ZSG and UCG)
are more non-trivial. Each variable in all these gauge conditions has a
unique gauge invariant combination.

An exception is the SG where even after fixing $\alpha \equiv 0$ in all
coordinates we have non-vanishing $\xi ^{t},$ where $\xi ^{t}(\mathbf{x})$
is the remnant gauge mode. From Eqs.\ (\ref{eq2}) and (\ref{eq4}), for $%
\delta p=c_{s}^{2}\delta \varrho $ with $c_{s}^{2}\equiv \dot{p}/\dot{\varrho%
}$, we have the solution $\delta \varrho \propto H(\varrho +p/c^{2})$ which
is exactly the behavior of the gauge mode as we have $\delta \widehat{%
\varrho }=\delta \varrho +c^{-1}3H(\varrho +p/c^{2})\xi ^{t}(\mathbf{x})$.
Thus, this solution can be \textit{removed} as a fictitious gauge mode. By
removing this gauge mode, and by \textit{setting} $\xi ^{t}(\mathbf{x})=0$
so $\delta =0$, the result becomes \textit{identical} to taking the UDG. The
complication caused by the remnant gauge mode in the SG can be overcome by
removing the gauge mode (as we just did and Lifshitz did in \cite%
{Lifshitz-1946}), by transforming to another gauge, or by constructing the
gauge-invariant combinations in the SG as Field and Shepley did in \cite%
{Field-Shepley-1968}. In any case the results are the same as our analysis
made above under the other fundamental gauge conditions: without any remnant
gauge mode.

In the zero-pressure fluid with general $K$ and $\Lambda$, the complete
solutions in all fundamental gauge conditions are presented in the Appendix %
\ref{App:MDE-solutions} for these two modes.


\subsection{Disagreements in the literature}

The original comments on the fictitious nature of $k^{2}=0$ and $3K$ modes
in spherical geometry were made by Lifshitz \cite{Lifshitz-1946,
Lifshitz-Khalatnikov-1963} and Bardeen \cite{Bardeen-1980}; see also \cite%
{Harrison-1967}. Here we analyse their arguments.

Lifshitz has introduced a scalar harmonic function $Q$ and constructed the
vector and tensor harmonic functions as (Eqs.\ (3.4), (3.10) and (3.11) in
\cite{Lifshitz-1946})
\begin{equation}
\Delta Q=-(n^{2}-1)Q,\quad P_{i}\equiv {\frac{1}{n^{2}-1}}Q_{,i},\quad
P_{ij}\equiv {\frac{1}{n^{2}-1}}Q_{,i|j}+{\frac{1}{3}}\gamma _{ij}Q.
\label{Lifshitz-harmonic}
\end{equation}%
Thus, $k^{2}=n^{2}-1$ with $n=1,2,\dots $; $P_{ij}$ is traceless with $%
P_{i}^{i}=0$. In a footnote below this equation Lifshitz mentions that $%
P_{ij}$ \textit{cannot} be constructed for $n=1$ and $2$; $P_{i}$ and $%
P_{ij} $ diverge (indeterminate as we have $Q_{,i}=0$ for $n=1$ with $%
Q=\sum_{\ell ,m}Q_{1\ell m}=Q_{100}=\mathrm{constant}$), and $P_{ij} $
vanishes for $n=2$ (we have $Q_{,i|j}=-\gamma _{ij}Q$ for $n=2$ with $%
Q=\sum_{\ell ,m}Q_{2\ell m}$); see Eq.\ (\ref{Q_100}). Thus, the vector and
tensor harmonic functions $P_{i}$ and $P_{ij}$ constructed in this way have
trouble in handling $n=1$ and $2$ modes properly.

Lifshitz has taken the synchronous gauge setting $\alpha \equiv 0\equiv
\beta $ in our notation; thus we have $\alpha =0$ and $\chi ={\frac{a^{2}}{c}%
}\dot{\gamma}$ in our equations. The spatial metric is expanded using the
tensor harmonic function as (Eq.\ (4.1) in \cite{Lifshitz-1946})
\begin{equation}
\widetilde{g}_{ij}\equiv a^{2}\left( \gamma _{ij}+h_{ij}\right) ,\quad
h_{ij}\equiv \mu {\frac{1}{3}}\gamma _{ij}Q+\lambda P_{ij},\quad h\equiv
h_{i}^{i}=\mu Q.  \label{Lifshitz-metric}
\end{equation}%
Compared with our metric convention in Eq.\ (\ref{metric}), we have
\begin{equation}
\varphi ={\frac{1}{6}}(\mu +\lambda )Q,\quad \gamma ={\frac{\lambda }{%
2(n^{2}-1)}}Q.
\end{equation}%
Notice that $\lambda $ is involved with a $1/(n^{2}-1)$ factor which is
troublesome for $n=1$.

Below Eq.\ (4.5) Lifshitz has made a major comment concerning our issue:
\textquotedblleft For $n=1,2$, we must put $\lambda =0$ because the tensor $%
P_{ij}$ does not exist for these values of $n$." By setting $\lambda =0$ we
have $\gamma =0$ in our notation, thus together with $\beta =0$ we have $%
\chi =0$. Thus, for these two modes Lifshitz sets $\alpha =0$ and $\chi =0$
simultaneously. As we set $\alpha =0$ and $\chi =0$ simultaneously, with $%
e=0=\Pi $, all variables in Eqs.\ (\ref{eq1})-(\ref{eq7}) vanish. Thus, all
perturbations disappear for these two modes $n=1$ and $2$. Setting $\alpha
=0=\chi $, however, is like imposing two temporal gauge (hypersurface,
slicing) conditions simultaneously which is \textit{not} allowed even for $%
n=1$ and $2$. Although $P_{i}$ and $P_{ij}$ cannot be constructed for $n=1$
and/or $2$ it is merely due to the way of constructing $P_{i}$ and $P_{ij}$
in Eq.\ (\ref{Lifshitz-harmonic}) which can be avoided by simply not
introducing such vector and tensor harmonics, see our Eq.\ (\ref{metric}).

Bardeen has similarly introduced harmonic functions as (Eqs.\ (2.7)-(2.9) in
\cite{Bardeen-1980}):
\begin{equation}
\Delta Q=-k^{2}Q,\quad Q_{i}\equiv -{\frac{1}{k}}Q_{,i},\quad Q_{ij}\equiv {%
\frac{1}{k^{2}}}Q_{,i|j}+{\frac{1}{3}}\gamma _{ij}Q.
\label{Bardeen-harmonic}
\end{equation}%
Compared with Lifshitz's notation we have $Q_{ij}=P_{ij}$ and $Q_{i}=-kP_{i}$%
; thus $Q_{i}$ and $Q_{ij}$ diverge (are indeterminate) for $n=1$, and $%
Q_{ij}$ vanishes for $n=2$. The metric tensor is expanded using the harmonic
functions as (Eq.\ (2.14) in \cite{Bardeen-1980}):
\begin{equation}
\widetilde{g}_{00}=-a^{2}(1+2AQ),\quad \widetilde{g}_{0i}=-a^{2}BQ_{i},\quad
\widetilde{g}_{ij}=a^{2}\left[ (1+2H_{L}Q)\gamma _{ij}+2H_{T}Q_{ij}\right] .
\label{Bardeen-metric}
\end{equation}%
Compared with our metric convention (which is in fact the convention of
Bardeen's other work  \cite{Bardeen-1988}), in Eq.\ (\ref{metric}), we have
\begin{equation}
\alpha =AQ,\quad \beta =-{\frac{1}{k}}BQ,\quad \varphi =\left( H_{L}+{\frac{1%
}{3}}H_{T}\right) Q,\quad \gamma ={\frac{1}{k^{2}}}H_{T}Q,
\end{equation}%
thus $B$ and $H_{T}$ are involved with $1/k$ and $1/k^{2}$ factors which are
troublesome for vanishing $k$; see below Eq.\ (211) in \cite{Ellis-van
Elst-1998}.

Concerning $n=1$ and $2$ modes, below Eq.\ (4.9) Bardeen mentions:
\textquotedblleft A spatially homogeneous perturbation or the lowest
inhomogeneous mode $k^{2}=3K$ in a closed universe require special treatment
in that $Q_{i}$ and/or $Q_{ij}$ vanish identically, $\Phi _{H}$, $\Phi _{A}$%
, and $v_{s}$ are no longer gauge-invariant $\dots $ A homogeneous scalar
perturbation is really no perturbation at all, but an inappropriate choice
of background." Compared with our notation, we have (Eqs.\ (3.9)-(3.11) in
\cite{Bardeen-1980}):
\begin{eqnarray}
&&\varphi _{\chi }\equiv \varphi -{\frac{H}{c}}\chi =\left[ H_{L}+{\frac{1}{3%
}}H_{T}+H\left( {\frac{a}{kc}}B-{\frac{a^{2}}{k^{2}c^{2}}}\dot{H}_{T}\right) %
\right] Q\equiv \Phi _{H}Q,  \notag \\
&&\alpha _{\chi }\equiv \alpha -{\frac{1}{c}}\dot{\chi}=\left[ A+{\frac{1}{kc%
}}\left( aB\right) ^{\displaystyle{\cdot }}-{\frac{1}{k^{2}c^{2}}}\left(
a^{2}\dot{H}_{T}\right) ^{\displaystyle{\cdot }}\right] Q\equiv \Phi _{A}Q,
\notag \\
&&v_{\chi }\equiv v-{\frac{c}{a}}\chi ={\frac{1}{k}}\left( v_{B}-{\frac{a}{k}%
}\dot{H}_{T}\right) Q\equiv {\frac{1}{k}}v_{s}Q,  \label{GI-Bardeen}
\end{eqnarray}%
where $v_{B}$ is Bardeen's $v$; compared with our $v$ we have
\begin{equation}
\widetilde{u}_{i}\equiv -{\frac{a}{c}}v_{,i},\quad {\frac{\widetilde{u}^{i}}{%
\widetilde{u}^{0}}}=\left( -{\frac{1}{c}}v+\beta \right) ^{|i}\equiv {\frac{1%
}{c}}v_{B}Q^{i}=-{\frac{1}{kc}}v_{B}Q^{|i},\quad \mathrm{thus},\quad v={%
\frac{1}{k}}\left( v_{B}-cB\right) Q.
\end{equation}%
Although Bardeen's definitions of $\Phi _{H}$, $\Phi _{A}$ and $v_{s}$ in
Eq.\ (\ref{GI-Bardeen}) have problems for $k=0$, in our notation, $\varphi
_{\chi }$, $\alpha _{\chi }$ and $v_{\chi }$, thus $\Phi _{H}$, $\Phi _{A}$
and $v_{s}$, are gauge invariant \textit{independently} of the value of $k$.

Concerning the homogeneous perturbation: the $n=1$ (homogeneous) mode scalar
perturbation is a correct result from the perturbation  and is handled
properly while maintaining the homogeneous and isotropic nature of the
background. This becomes clear in the Einstein static model as presented in
Eq.\ (\ref{varrho-solution-k=0}). Compared with this proper treatment, the
perturbation of the background equations, although it happens to give the
same result as in Eq.\ (\ref{BG-instability}), is a hand-waving procedure.
Rigorously, we have to perturb the original Einstein's equation around the
background, instead of simply perturbing the background equations, see
Section \ref{perturb-BG}.

Below Eq.\ (6.26) of \cite{Bardeen-1980} Bardeen addresses the $n=2$ mode by
analyzing it in the UEG, and states \textquotedblleft Since $Q_{ij}$
vanishes identically, Eq.\ (6.23) no longer applies." Bardeen's Eq.\ (6.23)
is a combination of our Eqs.\ (\ref{eq1}) and (\ref{eq3}) which do apply and
gives $v_{\kappa }=0$. Although many variables vanish in this gauge, there
are still surviving ones in the same gauge (these are $\varphi _{\kappa }$
and $\chi _{\kappa }$) as in Eq.\ (\ref{n=2-varphi-eq}). Concerning these
variables, below Eq.\ (6.27), Bardeen mentions \textquotedblleft The
amplitude [$\varphi _{\kappa }$] now depends on the way spatial coordinates
are propagated from one hypersurface to the next through the hypersurface
condition Eq.\ (5.22). The traceless part of the metric tensor perturbation
and the spatial curvature perturbation vanish. The absence of any physical
\textit{adiabatic} mode when $k^{2}=3K$ was first recognized by Lifshitz and
Khalatnikov."

Bardeen's Eq.\ (5.22) is our Eq.\ (\ref{eq1}). Together with Eq.\ (\ref{eq5}%
) it gives Eqs.\ (\ref{n=2-varphi-eq}) and (\ref{n=2-varphi-eq-2}). We find
no reason why these equations and behaviors of the gauge-invariant variables
$\varphi _{\kappa }$ and $\chi _{\kappa }$ should be regarded as coordinate
effects. In Bardeen's statement \textquotedblleft the traceless part of the
metric tensor perturbation" is $H_{T}$ in Eq.\ (\ref{Bardeen-metric}), thus
our $\gamma $, which can be set zero as the spatial gauge condition (without
any effect in our formulation as we are using $\chi $, which remains the
same). However, his \textquotedblleft spatial curvature perturbation" is our
$\varphi _{\kappa }$ which follows Eq.\ (\ref{n=2-varphi-eq}) and has
\textit{no} reason to vanish; if it vanishes $\chi _{\kappa }$ should vanish
as well which implies that all perturbation variables in the UEG vanish.
Vanishing $\varphi $ after imposing the UEG is like imposing two gauge
conditions\ (the UEG and the UCG) simultaneously, which is \textit{not}
allowed for general perturbation including the $k^{2}=3K$ mode.

%
%

\section{Stability of the Einstein static model with pressure}

\label{Einstein-static}

Einstein proposed in 1917 a static and closed world model by employing the
cosmological constant in a closed universe \cite{Einstein-1917}, for a
centennial review see \cite{ORaifeartaigh-2017}. Here, we consider the
presence of an additional pressure. In the static background, Eq.\ (\ref{BG}%
) gives field equations:
\begin{equation}
{\frac{8\pi G}{3}}\varrho _{0}={\frac{Kc^{2}}{a_{0}^{2}}}-{\frac{\Lambda
c^{2}}{3}},\quad 4\pi G\left( \varrho _{0}+{\frac{3p_{0}}{c^{2}}}\right)
=\Lambda c^{2},\quad 4\pi G\left( \varrho _{0}+{\frac{p_{0}}{c^{2}}}\right) =%
{\frac{Kc^{2}}{a_{0}^{2}}},  \label{BG-static}
\end{equation}%
thus
\begin{equation}
a_{0}^{2}={\frac{Kc^{2}}{4\pi G(1+w)\varrho _{0}}}={\frac{1+3w}{1+w}}{\frac{K%
}{\Lambda }}.  \label{BG-static-2}
\end{equation}%
We have $w>-{{1}/{3}}$ for $K>0$ and $\Lambda >0$.

To the linear order, ignoring stress, Eqs.\ (\ref{eq1})-(\ref{eq7}) give
\begin{eqnarray}
&&\kappa =-3\dot{\varphi}-c{\frac{\Delta }{a_{0}^{2}}}\chi ,
\label{eq1-static} \\
&&4\pi G\delta \varrho +c^{2}{\frac{\Delta +3K}{a_{0}^{2}}}\varphi =0,
\label{eq2-static} \\
&&\kappa +c{\frac{\Delta +3K}{a_{0}^{2}}}\chi -{\frac{12\pi G}{c^{2}}}%
a_{0}\left( \varrho _{0}+{\frac{p_{0}}{c^{2}}}\right) v=0,
\label{eq3-static} \\
&&\dot{\kappa}+c^{2}{\frac{\Delta }{a_{0}^{2}}}\alpha =4\pi G\left( \delta
\varrho +{\frac{3\delta p}{c^{2}}}\right) ,  \label{eq4-static} \\
&&\varphi +\alpha -{\frac{1}{c}}\dot{\chi}=0,  \label{eq5-static} \\
&&\delta \dot{\varrho}-\left( \varrho _{0}+{\frac{p_{0}}{c^{2}}}\right)
\left( \kappa +{\frac{\Delta }{a_{0}}}v\right) =0,  \label{eq6-static} \\
&&\dot{v}={\frac{1}{a_{0}}}\left( {\frac{\delta p}{\varrho _{0}+{\frac{p_{0}%
}{c^{2}}}}}+c^{2}\alpha \right) .  \label{eq7-static}
\end{eqnarray}%
The gauge transformation properties in Eq.\ (\ref{GT}) become
\begin{eqnarray}
&&\widehat{\alpha }=\alpha -{\frac{1}{c}}\dot{\xi}^{t},\quad \widehat{%
\varphi }=\varphi ,\quad \widehat{\chi }=\chi -\xi ^{t},\quad \widehat{%
\kappa }=\kappa +c{\frac{\Delta }{a_{0}^{2}}}\xi ^{t},\quad \delta \widehat{%
\varrho }=\delta \varrho ,\quad \delta \widehat{p}=\delta p,\quad \widehat{v}%
=v-{\frac{c}{a_{0}}}\xi ^{t},  \notag \\
&&\widehat{\Pi }=\Pi ,\quad \delta \widehat{\phi }=\delta \phi -{\frac{1}{c}}%
\dot{\phi}_{0}\xi ^{t}.  \label{GT-static}
\end{eqnarray}%
Thus, $\delta \varrho $ and $\varphi $ are naturally gauge-invariant, and $%
\kappa $ becomes gauge-invariant for $\Delta =0$. As we have $\Phi =0$ we
cannot use the equations and solutions in Sections \ref{sec:Phi-eqs} and \ref%
{MDE-solutions}.

\textit{Without} imposing the gauge condition, from Eqs.\ (\ref{eq4-static}%
), (\ref{eq6-static}) and (\ref{eq7-static}), we can derive (except for $%
\Delta +3K=0$ case, where we have $\delta \varrho =0$ from Eq.\ (\ref%
{eq2-static}), see below) the following second-order density perturbation
equation:
\begin{equation}
\delta \ddot{\varrho}=\left[ 4\pi G(1+w)(1+3c_{s}^{2})\varrho
_{0}+c_{s}^{2}c^{2}{\frac{\Delta }{a_{0}^{2}}}\right] \delta \varrho ={\frac{%
c^{2}}{a_{0}^{2}}}\left[ K+c_{s}^{2}\left( \Delta +3K\right) \right] \delta
\varrho .  \label{delta-eq-static}
\end{equation}%
The solution is (Eq.\ (226) in \cite{Harrison-1967} and Eq.\ (34) in \cite%
{Gibbons-1987})
\begin{equation}
\delta \varrho \propto \varphi \propto e^{\pm \sqrt{4\pi
G(1+w)(1+3c_{s}^{2})\varrho _{0}-c_{s}^{2}c^{2}{k^{2}/a_{0}^{2}}}t}=e^{\pm
\sqrt{K-c_{s}^{2}(k^{2}-3K)}ct/a_{0}}\propto e^{\pm \sqrt{\lbrack
1-(L+3)(L-1)c_{s}^{2}]K}ct/a_{0}},  \label{varphi-solution}
\end{equation}%
where we set
\begin{equation}
\Delta =-k^{2}=-(n^{2}-1)K=-L(L+2)K,\quad (L=n-1=0,1,2,\dots ).
\end{equation}%
For a stable solution we need
\begin{equation}
c_{s}^{2}\geq {\frac{1}{(L+3)(L-1)}},  \label{stability-criterion}
\end{equation}%
and for $L\geq 2$ we have $c_{s}^{2}\geq {1/5}$ \cite{Harrison-1967,
Gibbons-1987, Gibbons-1988, Barrow-etal-2003}. The $L=0$ and $1$ modes are
always unstable; although these two modes were often ignored
in the literature as fictitious perturbations, we have shown that these are physical.

Equation (\ref{delta-eq-static}) is \textit{not} applicable for $L=1$, thus $%
k^{2}=3K$. In this case from Eq.\ (\ref{eq2-static}) we have $\delta \varrho
=0$, but as $\varphi $ obeys the same equation, the solution in terms of $%
\varphi $ remains valid, see Section \ref{sec:k2=3K}.

\emph{\ }

\subsection{Complete solutions for general $k^2 = L (L + 2) K$}

\label{sec:general-mode}

The variables $\delta $ and $\varphi $ are gauge invariant, and the
solutions for them are given in Eq.\ (\ref{varphi-solution}). Solutions for
other variables in all other gauges can be derived from a known solution.
From Eqs.\ (\ref{eq1-static})-(\ref{eq7-static}), or using the gauge
transformation properties in Eq.\ (\ref{GT-static}), for the relatively
growing mode, we can show
\begin{eqnarray}
&&\varphi \propto e^{+\sqrt{[1-(L+3)(L-1)c_{s}^{2}]K}ct/a_{0}},\quad \delta
=(1+w)(L+3)(L-1)\varphi ;\quad \kappa _{\chi }=-3\sqrt{%
[1-(L+3)(L-1)c_{s}^{2}]K}{\frac{c}{a_{0}}}\varphi ,  \notag \\
&&v_{\chi }=-c\sqrt{{\frac{1-(L+3)(L-1)c_{s}^{2}}{K}}}\varphi ,\quad \alpha
_{\chi }=-\varphi ;\quad \kappa _{v}=(L+3)(L-1)\sqrt{[1-(L+3)(L-1)c_{s}^{2}]K%
}{\frac{c}{a_{0}}}\varphi  \notag \\
&&\chi _{v}=a_{0}\sqrt{{\frac{1-(L+3)(L-1)c_{s}^{2}}{K}}}\varphi ,\quad
\alpha _{v}=-(L+3)(L-1)c_{s}^{2}\varphi ;\quad \chi _{\kappa }=3a_{0}{\frac{%
\sqrt{1-(L+3)(L-1)c_{s}^{2}}}{L(L+2)\sqrt{K}}}\varphi ,  \notag \\
&&v_{\kappa }=-c{\frac{(L+3)(L-1)}{L(L+2)\sqrt{K}}}\sqrt{%
1-(L+3)(L-1)c_{s}^{2}}\varphi ,\quad \alpha _{\kappa }=-{\frac{(L+3)(L-1)}{%
L(L+2)}}(1+3c_{s}^{2})\varphi .  \label{general-solutions}
\end{eqnarray}%
These are complete solutions. Since all these variables are proportional to $%
\varphi $, they share the same equation in (\ref{delta-eq-static}).

The solutions for $v_{\kappa }$, $\alpha _{\kappa }$ and $\chi _{\kappa }$
diverge when $L=0$. This happens because we \textit{cannot} take the UEG for
$L=0$ and $\kappa $ becomes gauge invariant for this situation. As we have $%
\varphi =\sum_{L,\ell ,m}\varphi _{L\ell m}Q_{L\ell m}$ (see Eq.\ (\ref%
{Harmonic-expansion}) for a proper expansion), the breakdown of the UEG for $%
L=0$ implies the breakdown of the UEG in general in Einstein static model.
Thus the UEG is \textit{not} available for the Einstein static model.
Excluding solutions in the UEG, the above solutions are generally \textit{%
valid} for all $L$ including $L=0$ and $1$; the latter special cases will be
displayed below.


\subsection{$k^2 = 3K$ mode ($L = 1$)}

\label{sec:k2=3K}

From Eq.\ (\ref{eq2-static}) we have $\delta \varrho =0$ which is naturally
gauge invariant. From Eqs.\ (\ref{eq1-static}), (\ref{eq4-static}) and (\ref%
{eq5-static}) another naturally gauge invariant variable $\varphi $ follows:
\begin{equation}
\ddot{\varphi}={\frac{Kc^{2}}{a_{0}^{2}}}\varphi =4\pi G\left( 1+w\right)
\varrho _{0}\varphi ,\quad \mathrm{thus,}\quad \varphi \propto e^{\pm \sqrt{K%
}ct/a_{0}}\propto e^{\pm \sqrt{4\pi G(1+w)\varrho _{0}}t},
\label{k2=3K-solution}
\end{equation}%
which is exponentially unstable; although $\delta \varrho =0$, the solution
for $\varphi $ in Eq.\ (\ref{varphi-solution}) remains valid. By taking
either the UEG ($\kappa \equiv 0$) or the CG ($v\equiv 0$) we have $\kappa
=v=\alpha =0$, thus
\begin{equation}
v_{\kappa }=0=\alpha _{\kappa },\quad \kappa _{v}=0=\alpha _{v}.
\end{equation}%
In both gauges we still have non-trivial equations in (\ref{eq1-static}) and
(\ref{eq5-static})
\begin{equation}
\dot{\varphi}={\frac{Kc}{a_{0}^{2}}}\chi ,\quad \varphi ={\frac{1}{c}}\dot{%
\chi},
\end{equation}%
a combination of which leads to Eq.\ (\ref{k2=3K-solution}). For relatively
growing solutions we can show,
\begin{equation}
\varphi \propto e^{+\sqrt{K}ct/a_{0}},\quad \delta =0,\quad v_{\chi }=-{%
\frac{c}{\sqrt{K}}}\varphi ,\quad \kappa _{\chi }=-3{\frac{\sqrt{K}c}{a_{0}}}%
\varphi ,\quad \alpha _{\chi }=-\varphi ,\quad \chi _{v}=\chi _{\kappa }={%
\frac{a_{0}}{\sqrt{K}}}\varphi ,  \label{solutions-k=3}
\end{equation}%
so the general solutions in Eq.\ (\ref{general-solutions}) remain \textit{%
valid} for $L=1$.


\subsection{$k^2 = 0$ mode ($L = 0$)}

\label{ES-k=0}

Equation (\ref{GT-static}) shows that besides $\delta \varrho $ and $\varphi
$, the variable $\kappa $ is also gauge invariant. Equation (\ref%
{delta-eq-static}) gives
\begin{equation}
\delta \ddot{\varrho}=4\pi G(1+w)(1+3c_{s}^{2})\varrho _{0}\delta \varrho
=\left( 1+3c_{s}^{2}\right) {\frac{Kc^{2}}{a_{0}^{2}}}\delta \varrho ,
\label{delta-eq-static-k=0}
\end{equation}%
with the solution
\begin{equation}
\delta \varrho \propto e^{\pm \sqrt{4\pi G(1+w)(1+3c_{s}^{2})\varrho _{0}}%
t}=e^{\pm \sqrt{(1+3c_{s}^{2})K}ct/a_{0}}.  \label{varrho-solution-k=0}
\end{equation}%
This mode is exponentially unstable for $c_{s}^{2}>-{1/3}$. All
surviving variables satisfy the same equation as $\delta \varrho $ in Eq.\ (%
\ref{delta-eq-static-k=0}). For relatively growing modes we have
\begin{eqnarray}
&&\varphi \propto e^{+\sqrt{(1+3c_{s}^{2})K}ct/a_{0}},\quad \delta =-3\left(
1+w\right) \varphi ,\quad \kappa =-3\sqrt{(1+3c_{s}^{2})K}{\frac{c}{a_{0}}}%
\varphi ,\quad \alpha _{v}=3c_{s}^{2}\varphi ,\quad \alpha _{\chi }=-\varphi
,  \notag \\
&&v_{\chi }=-c\sqrt{\frac{1+3c_{s}^{2}}{K}}\varphi ,\quad \chi _{v}=a_{0}%
\sqrt{\frac{1+3c_{s}^{2}}{K}}\varphi .  \label{solutions-k=0}
\end{eqnarray}%
Thus, the general solutions in Eq.\ (\ref{general-solutions}) remain \textit{%
valid} for $L=0$. Since $\kappa $ is gauge invariant, we do not have
variables like $v_{\kappa }$, $\alpha _{\kappa }$ and $\chi _{\kappa }$
which display divergent behaviour in Eq.\ (\ref{general-solutions}).


\subsection{Stability of the static background}

\label{perturb-BG}

By directly perturbing the background equations to the linear order, with $%
a=a_{0}+\delta a$, $\varrho =\varrho _{0}+\delta \varrho $ and $%
p=p_{0}+\delta p$, the first two in equation (\ref{BG}) give
\begin{equation}
{\frac{\delta \ddot{a}}{a_{0}}}=-{\frac{4\pi G}{3}}\left( \delta \varrho +{%
\frac{3\delta p}{c^{2}}}\right) ,\quad {\frac{4\pi G}{3}}\delta \varrho =-{%
\frac{Kc^{2}}{a_{0}^{2}}}{\frac{\delta a}{a_{0}}},\quad \mathrm{thus,}\quad
\delta \ddot{a}=\left( 1+3c_{s}^{2}\right) {\frac{Kc^{2}}{a_{0}^{2}}}\delta
a,
\end{equation}%
with the solution (Eq.\ (225) in \cite{Harrison-1967}):
\begin{equation}
\delta a\propto \delta \varrho \propto e^{\pm \sqrt{(1+3c_{s}^{2})K}%
ct/a_{0}}.  \label{BG-instability}
\end{equation}%
Thus, the static background is exponentially unstable for $c_{s}^{2}>-{1/3}$.

The behavior of $\delta \varrho $ happens to \textit{coincide} with the
homogeneous perturbation mode in Eq.\ (\ref{varrho-solution-k=0}). We note,
however, that in our proper perturbation of Einstein equations with the
results in Eq.\ (\ref{solutions-k=0}), even for the homogeneous mode,
besides $\delta g_{ij}$, the quantity $\delta g_{00}$ is excited in the ZSG (%
$\delta g_{0i}\equiv 0$), and both $\delta g_{00}$ and $\delta g_{0i}$ are
excited in the CG. By perturbing the background metric with $a=a_{0}+\delta a
$ we effectively have $\alpha =\delta a/a_{0}=\varphi $ which \textit{differs%
} from our result in the ZSG with $\alpha =-\varphi $, and in the CG with $%
\alpha =3c_{s}^{2}\varphi $. Thus, despite a coincidence in $\delta \varrho $%
, perturbing the Friedmann equation alone \cite{Eddington-1930} is a
hand-waving procedure. The correct way is to perturb the original Einstein
equations, as we did in Section \ref{ES-k=0}; see Section 5.7 in \cite%
{Harrison-1967}.


\subsection{Einstein static background with scalar field}

In the Einstein static background, we set $\ddot{\phi}_{0}\equiv 0$ but $%
\dot{\phi}_{0}\neq 0$ \cite{Barrow-etal-2003}; if $\dot{\phi}_{0}=0$ we have
$w=-1$, $\delta \varrho =0=\delta p$ and the case becomes trivial. From Eq.\
(\ref{BG-EOM}) we have $V_{,\phi }=0$ and so $V=V_{0}$. Thus, we have
\begin{equation}
\varrho _{0}={\frac{1}{2}}\dot{\phi}_{0}^{2}+V_{0},\quad p_{0}={\frac{1}{2}}%
\dot{\phi}_{0}^{2}-V_{0},  \label{BG-EOM-static}
\end{equation}%
for the background, and
\begin{eqnarray}
&&\delta \varrho =\delta p=\dot{\phi}_{0}\delta \dot{\phi}-\dot{\phi}%
_{0}^{2}\alpha ,\quad \left( \varrho _{0}+p_{0}\right) v={\frac{1}{a_{0}}}%
\dot{\phi}_{0}\delta \phi ,  \label{pert-fluid-MSF-static} \\
&&\delta \ddot{\phi}-{\frac{\Delta }{a^{2}}}\delta \phi =\dot{\phi}%
_{0}\left( \kappa +\dot{\alpha}\right) ,  \label{pert-EOM-static}
\end{eqnarray}%
for the perturbation. Equations (\ref{BG-static}) and (\ref{BG-static-2})
for the background and Eqs.\ (\ref{eq1-static})-(\ref{eq7-static}) for the
perturbation remain valid with the fluid quantities replaced as above.

Equations (\ref{BG-static}) and (\ref{BG-static-2}) give
\begin{equation}
{\frac{8\pi G}{3}}\varrho _{0}={\frac{8\pi G}{3}}\left( {\frac{1}{2}}\dot{%
\phi}_{0}^{2}+V_{0}\right) ={\frac{K}{a_{0}^{2}}}-{\frac{\Lambda }{3}},\quad
4\pi G\left( \varrho _{0}+3p_{0}\right) =8\pi G\left( \dot{\phi}%
_{0}^{2}-V_{0}\right) =\Lambda ,\quad 4\pi G\left( \varrho _{0}+p_{0}\right)
=4\pi G\dot{\phi}_{0}^{2}={\frac{K}{a_{0}^{2}}}.  \label{BG-static-MSF}
\end{equation}%
a. In the presence of $\Lambda $, we have
\begin{equation}
{\frac{K}{a_{0}^{2}}}=\Lambda {\frac{1+w}{1+3w}}={\frac{\Lambda }{2}}{\frac{%
\dot{\phi}_{0}^{2}}{\dot{\phi}_{0}^{2}-V_{0}}},\quad w\equiv {\frac{p_{0}}{%
\varrho _{0}}}={\frac{\dot{\phi}_{0}^{2}-2V}{\dot{\phi}_{0}^{2}+2V}},
\end{equation}%
thus, for $K>0$ and $\Lambda >0$ we have $\dot{\phi}_{0}^{2}>V_{0}$ and $-{%
{1}/{3}}<w<1$.

\noindent b. In the absence of $\Lambda $, we have
\begin{equation}
\dot{\phi}_{0}^{2}=V_{0}={\frac{K}{4\pi Ga_{0}^{2}}},\quad w=-{\frac{1}{3}}.
   \label{MSF-case-b}
\end{equation}

As we have $\delta p=\delta \varrho $ the rest of the analyses made for the
fluid case remains valid with $c_{s}^{2}\equiv \delta p/\delta \varrho =1$;
notice that, as we have $V_{,\phi }=0$ for the background, see Eqs.\ (\ref%
{BG-EOM}) and (\ref{pert-fluid-MSF}), this is true \textit{independently} of
the scalar field potential. Thus, {\it all $L\geq 2$ modes are stable,
but $L=0$ and $1$ modes are unstable}, as in the fluid case.
The equation of motion in
Eq.\ (\ref{pert-EOM-static}) is consistent with Eqs.\ (\ref{eq6-static}) and
(\ref{eq7-static}), and $\delta \phi $ is determined by $\delta \phi =a_{0}%
\dot{\phi}_{0}v$.

%
%

\section{The Taub constraint}

\label{Taub-constraint}

Losic and Unruh \cite{Losic-Unruh-2005} state in their conclusions that
\textquotedblleft the requirement that the second order Einstein constraint
equations be integrable demands that any inhomogenous linear mode
perturbations of the Einstein static universe must be accompanied by the
homogenous linear mode with comparable amplitude."
In order for a solution of the linearized equation to be a proper solution
of the exact equation, it should satisfy a constraint to the second order.
We may call it the Taub constraint \cite{Taub-1961}. Here we evaluate the
Taub constraint based on a timelike Killing vector in an Einstein static
background. The Taub constraint is derived in the Appendix \ref%
{App:Taub-constraint} considering the general background metric,
see Eqs.\ (\ref{Taub}) and (\ref{Taub-2}). Our result confirms Losic and Unruh's
above conclusion, but only for $c_{s}^{2}=0,$ whereas they were claiming it
was true for arbitrary $c_{s}^{2}$.


\subsection{Spherical harmonic expansion}

When we consider the spherical background geometry we need harmonics in
spherical geometry. We expand
\begin{eqnarray}
&&\varphi (t,\chi ,\theta ,\phi )\equiv \varphi _{\mathbf{k}}(t)Q(\mathbf{k}%
;\chi ,\theta ,\phi )\equiv \sum_{n,\ell ,m}\varphi _{n\ell m}(t)Q_{n\ell
m}(\chi ,\theta ,\phi )\equiv \sum_{n=1,2,\dots }\sum_{\ell
=0}^{n-1}\sum_{m=-\ell }^{\ell }\varphi _{n\ell m}(t)\Pi _{n\ell }(\chi
)Y_{\ell }^{m}(\theta ,\phi ),  \notag \\
&&\varphi _{n\ell m}(t)=\int \varphi (t,\chi ,\theta ,\phi )\Pi _{n\ell
}(\chi )Y_{\ell }^{m\ast }(\theta ,\phi )\sqrt{\gamma }d^{3}x,
\label{Harmonic-expansion}
\end{eqnarray}%
where $\Pi _{n\ell }(\chi )$ is an associated Legendre function with proper
normalization \cite{Harrison-1967}:
\begin{equation}
\int_{0}^{\pi }\Pi _{n\ell }(\chi )\Pi _{n^{\prime }\ell }(\chi )\sin ^{2}{%
\chi }d\chi =\delta _{nn^{\prime }},\quad \Pi _{n\ell }(\chi )=\sqrt{\frac{%
n\Gamma (n+\ell +1)}{\Gamma (n-\ell )}}{\frac{1}{\sqrt{\sin {\chi }}}}%
P_{n-1/2}^{-\ell -1/2}(\cos {\chi }).
\end{equation}%
Using the spatial metric in Eq.\ (\ref{spatial-metric}) it is convenient to
have
\begin{eqnarray}
&&\gamma _{\chi \chi }=1,\quad \gamma _{\theta \theta }=\sin ^{2}{\chi }%
,\quad \gamma _{\phi \phi }=\sin ^{2}{\chi }\sin ^{2}{\theta };\quad \gamma
^{\chi \chi }=1,\quad \gamma ^{\theta \theta }={\frac{1}{\sin ^{2}{\chi }}}%
,\quad \gamma ^{\phi \phi }={\frac{1}{\sin ^{2}{\chi }\sin ^{2}{\theta }}};
\notag \\
&&\Gamma _{\;\;\;\;\;\theta \theta }^{(\gamma )\chi }=-\sin {\chi }\cos {%
\chi },\quad \Gamma _{\;\;\;\;\;\phi \phi }^{(\gamma )\chi }=-\sin {\chi }%
\cos {\chi }\sin ^{2}{\theta },\quad \Gamma _{\;\;\;\;\;\chi \theta
}^{(\gamma )\theta }={\frac{\cos {\chi }}{\sin {\chi }}},\quad \Gamma
_{\;\;\;\;\;\phi \phi }^{(\gamma )\theta }=-\sin {\theta }\cos {\theta },
\notag \\
&&\Gamma _{\;\;\;\;\;\chi \phi }^{(\gamma )\phi }={\frac{\cos {\chi }}{\sin {%
\chi }}},\quad \Gamma _{\;\;\;\;\;\theta \phi }^{(\gamma )\phi }={\frac{\cos
{\theta }}{\sin {\theta }}},\quad \text{or }0\;\;\mathrm{otherwise}.
\end{eqnarray}

For $n = 1$ and $2$ the harmonic functions $Q$ are
\bea
   & & Q |_{n = 1}
       = Q_{100} = {1 \over \sqrt{2} \pi},
       \quad
       Q |_{n = 2}
       = Q_{200} + Q_{21-1} + Q_{210} + Q_{211}
       = {\sqrt{2} \over \pi} \left[ \cos{\chi}
       + \sin{\chi} \left( \cos{\theta}
       - 2 i \sin{\theta} \sin{\phi} \right) \right].
   \label{Q_100}
\eea
We have $Q_{,i} = 0$ for $n = 1$, and $Q_{,i|j} = - \gamma_{ij} Q$ for $n = 2$.

Using $n=L+1$, thus $k^{2}=n^{2}-1=L(L+2)$, we have
\begin{eqnarray}
&&\varphi (t,\mathbf{x})=\varphi (t,\chi ,\theta ,\phi )=\sum_{L=0,1,\dots
}\sum_{\ell =0}^{L}\sum_{m=-\ell }^{\ell }\varphi _{L\ell m}(t)\Pi _{L\ell
}(\chi )Y_{\ell }^{m}(\theta ,\phi ),  \notag \\
&&\Delta \varphi (t,\mathbf{x})=-\sum_{n,\ell ,m}(n^{2}-1)\varphi _{n\ell
m}(t)\Pi _{n\ell }Y_{\ell }^{m}=-\sum_{L,\ell ,m}L(L+2)\varphi _{L\ell
m}(t)\Pi _{L\ell }Y_{\ell }^{m}.  \label{varphi_nlm}
\end{eqnarray}%
It is convenient to have
\begin{eqnarray}
&&\int \varphi ^{2}\sqrt{\gamma }d^{3}x=\sum_{L,\ell ,m}|\varphi _{n\ell
m}|^{2},\quad \int \varphi ^{|i}\varphi _{,i}\sqrt{\gamma }d^{3}x=-\int
\varphi \Delta \varphi \sqrt{\gamma }d^{3}x=\sum_{L,\ell ,m}L(L+2)|\varphi
_{L\ell m}|^{2},  \notag \\
&&\int \sqrt{\gamma }\varphi ^{|ij}\varphi _{,i|j}d^{3}x=\int \varphi \Delta
(\Delta +2K)\varphi \sqrt{\gamma }d^{3}x=\sum_{L,\ell ,m}\left(
L^{2}+2L-2\right) L\left( L+2\right) K^{2}|\varphi _{n\ell m}|^{2}.
\end{eqnarray}


\subsection{The Taub constraint in two gauges}

Complete solutions are presented in Eq.\ (\ref{general-solutions}); $\varphi
$ and $\delta $ are gauge invariant and solutions in the UEG are not valid.
Using the mode expansion in Eq.\ (\ref{varphi_nlm}), the solutions can be
written as
\begin{eqnarray}
&&\varphi =\sum_{L,\ell ,m}\varphi _{L\ell m}(t)\Pi _{L\ell }Y_{\ell
}^{m}=\sum_{L,\ell ,m}e^{\sqrt{[1-(L+3)(L-1)c_{s}^{2}]K}c(t-t_{0})/a_{0}}%
\varphi _{L\ell m}(t_{0})\Pi _{L\ell }Y_{\ell }^{m};  \notag \\
&&\delta _{L\ell m}=(1+w)(L+3)(L-1)\varphi _{L\ell m};\quad \kappa _{\chi
L\ell m}=-3\sqrt{[1-(L+3)(L-1)c_{s}^{2}]K}{\frac{c}{a_{0}}}\varphi _{L\ell
m},  \notag \\
&&v_{\chi L\ell m}=-c\sqrt{{\frac{1-(L+3)(L-1)c_{s}^{2}}{K}}}\varphi _{L\ell
m},\quad \alpha _{\chi L\ell m}=-\varphi _{L\ell m};\quad \chi _{vL\ell
m}=a_{0}\sqrt{{\frac{1-(L+3)(L-1)c_{s}^{2}}{K}}}\varphi _{L\ell m},  \notag
\\
&&\kappa _{vL\ell m}=(L+3)(L-1)\sqrt{[1-(L+3)(L-1)c_{s}^{2}]K}{\frac{c}{a_{0}%
}}\varphi _{L\ell m},\quad \alpha _{vL\ell m}=-c_{s}^{2}(L+3)(L-1)\varphi
_{L\ell m}.  \label{linear-solutions-Taub}
\end{eqnarray}%
We have shown that these solutions are valid for all $L$ values.
For the $L = 0$ mode, the temporal dependence of all variables is
$\propto e^{\sqrt{(1 + 3 c_s^2)K} ct/a_0}$, and so is unstable for
$c_s^2 > - {1 / 3}$. For the $L = 1$ mode the temporal dependence of
all variables becomes $\propto e^{\sqrt{K} ct/a_0}$,  and so is unstable,
independently of $c_s^2$. For $c_s^2 > {1 / 5}$ all modes with
$L \ge 2$ become stable as presented in Eq.\ (\ref{stability-criterion})
\cite{Harrison-1967,
Gibbons-1987, Gibbons-1988, Barrow-etal-2003}.

Now, we can evaluate the Taub constraint in Eq.\ (\ref{Taub-cosmology-3})
using Eq.\ (\ref{eq2-E0-scalar}) and the linear solutions in Eq.\ (\ref%
{linear-solutions-Taub}). In the ZSG and the CG, respectively, we have
\begin{eqnarray}
&&\mathcal{T}_{\mathrm{ZSG}}=\sum_{L=0,1,\dots }\sum_{\ell
=0}^{L}\sum_{m=-\ell }^{\ell }\left[ 7L^{2}+14L-15-\left( 2L^{2}+4L-3\right)
(L+3)(L-1)c_{s}^{2}\right] K|\varphi _{L\ell m}|^{2}=0,  \label{Taub-ES-ZSG}
\\
&&\mathcal{T}_{\mathrm{CG}}=\sum_{L=0,1,\dots }\sum_{\ell
=0}^{L}\sum_{m=-\ell }^{\ell }\left[ 3\left( 2L^{2}+4L-5\right)
-(L+3)^{2}(L-1)^{2}c_{s}^{2}\right] K|\varphi _{L\ell m}|^{2}=0.
\label{Taub-ES-CG}
\end{eqnarray}%
The Taub constraint apparently depends on the gauge condition. Taub
constraint on the ZSG applies to the linear solutions in the ZSG, and
likewise for the CG. We have $\mathcal{T}_{L=1}>0$ and $\mathcal{T}%
_{L=0}<0$ in both gauges.

For $c_{s}^{2}=0$, in both gauges we have $\mathcal{T}_{\mathrm{L\geq 1}%
}>0$ and $\mathcal{T}_{\mathrm{L=0}}<0$, thus the Taub
constraint demands that the homogeneous perturbation ($L=0$ mode) should be
excited as long as we have non-vanishing inhomogeneous perturbation; the
solutions in Eq.\ (\ref{linear-solutions-Taub}) show that for $c_{s}^{2}=0$,
all perturbations are unstable proportional to $e^{\sqrt{K}ct/a_{0}}$
independently of $L$.

Similarly, in the ZSG, for $c_{s}^{2}>{{41}/{65}}$ we have $\mathcal{T}_{%
\mathrm{L=1}}>0$ while all contributions from the other modes are pure
negative. Thus, in this case, the $L=1$ mode should be excited as long as we
have any other perturbation mode in the ZSG. Concerning the $L=1$ mode, a
similar conclusion does not follow in the CG, as we need $c_{s}^{2}\leq 1$.

%
%
%

\section{Discussion}

\label{Discussion}

We have shown that the lowest two ($L = 0$ and $1$) perturbation modes
in closed universes are not fictitious perturbations,
see Secs.\ \ref{Linear-perturbation} and \ref{Physical-nature}.
However, the case is more subtle than normally considered as we often
have linearization stability
issues in a closed space with the Killing symmetry. The Einstein static model
is such a closed space with a timelike Killing vector. The Taub constraint
provides a constraint on quadratic combinations of linear order variables
for linearization stability to hold. We have derived the Taub constraint
in general background metric in the Appendix \ref{App:Taub-constraint};
these are Eqs.\ (\ref{Taub}) and (\ref{Taub-2}) for general background,
and Eqs.\ (\ref{Taub-cosmology-2}) and (\ref{Taub-cosmology-3})
for cosmological background. We have evaluated the Taub constraint in
the Einstein static model with a timelike Killing vector,
see Secs.\ \ref{Einstein-static} and \ref{Taub-constraint}.
The results are presented in Eqs.\ (\ref{Taub-ES-ZSG}) and (\ref{Taub-ES-CG})
for two fundamental gauge conditions available in the Einstein static
model with pressure. The result can be compared with other works as follows.

According to Losic and Unruh \cite{Losic-Unruh-2005} $\mathcal{T}_{\mathrm{%
L\geq 2}}$ should be pure positive; they ignored the $L=1$ mode as a
gauge mode. But in such a case the Taub constraint \textit{demands} the
presence of the $L=0$ mode which is negative. Although Losic and Unruh have
claimed this is the case for general $c_{s}^{2}$ , Eqs.\ (\ref{Taub-ES-ZSG})
and (\ref{Taub-ES-CG}) show that this is true only for $c_{s}^{2}=0$ in both
the ZSG and the CG. The gauge condition adopted by Losic and
Unruh, and whether they were using the same fluid as ours are
unclear to us.

For $c_{s}^{2}>{1/5}$ we have that the $L\geq 2$ modes are stable,
while $L=0$ and $1$ modes are unstable. Although Losic and Unruh have stated
that the perturbation should accompany the unstable $L=0$ mode, our result
does not confirm the case at least in our two gauge conditions; it is true
only for $c_{s}^{2}=0$ and in this case all modes are unstable.

Studying a conformal variation Gibbons \cite{Gibbons-1987, Gibbons-1988}
concluded that $\mathcal{T}_{\mathrm{L\geq 2}}>0$, $\mathcal{T}_{\mathrm{L=1}%
}=0$ and $\mathcal{T}_{\mathrm{L=0}}<0$ for $c_{s}^{2}=0$. Although we
expressed Gibbons' result using $\mathcal{T}$, his method is based on
second-order variation of entropy and the exact relation to our method is
unclear. The conformal variation, $\delta g_{ab}=\phi g_{ab}$, implies $%
\alpha =\varphi $ and $\chi =0$ in our notation, and this \textit{differs}
from the ZSG where $\alpha =-\varphi $ and $\chi =0$. The conformal
variation is not available in a proper perturbation theory.

The presence of stable perturbation modes with $L\geq 2$ for $c_{s}^{2}$ $%
>1/5$ has suggested the Einstein static model with pressure might be a
potential evolutionary stage in the early universe, before inflation
without singularity \cite{Barrow-etal-2003}, \cite{Ellis-Maartens-2004}.
An Einstein static phase supported by a massless scalar field belongs to this
case with $c_s^2 = 1$, see below Eq.\ (\ref{MSF-case-b}).
Although it has been suggested that the excitation of $L\geq 2$ modes should
accompany the homogeneous ($L=0$) mode \cite{Losic-Unruh-2005}, which is
always unstable, our result shows that this applies only for $c_{s}^{2}=0$%
. For $c_{s}^{2}>1/5$, the Taub constraint in two gauge conditions in Eqs.\
(\ref{Taub-ES-ZSG}) and (\ref{Taub-ES-CG}) shows that it is not necessary to
accompany $L=0$ and/or $L=1$ modes, both of which are unstable;
for $c_{s}^{2}=1/5$, we can show that the $L=0$ mode is
negative, the $L=1$ to $3$ modes are positive and the $L\geq 4$ modes are
negative again in the ZSG, whereas the $L=0$ mode is negative; the $L=$ $1-4$
modes are positive and the $L\geq 5$ modes are negative again in the CG.
As both the $L=0$ and $L=1$ modes are unstable even for $c_{s}^{2}>1/5$,
these two modes must be suppressed to have a stable Einstein static stage.
How to avoid exciting these lowest two modes for a successful realization is
a question yet to be answered.

Previously the $L = 0$ and $1$ (thus $n = 1$ and $2$) modes were generally regarded as fictitious, thus largely ignored in the literature, see \cite{White-Scott-1996, Lewis-Challinor-Lasenby-2000, Lesgourgues-Tram-2014}.
The physical nature of these two modes with newly restored honor in this paper implies that one needs to properly take into account of these two modes in the future cosmological calculation of closed Friedmann world model.
These include the full sky galaxy correlation function and power spectrum (see \cite{Tansella-Bonvin-etal-2018, Tansella-etal-2018} in flat background),
the CMB (cosmic microwave background radiation) anisotropy power spectra (see \cite{Planck-2014, Ivanov-etal-2020} for contributions of monopole and dipole in flat background),
and others like the luminosity distance and redshift (see \cite{Biern-Yoo-2017, Fanizza-Yoo-Biern-2018} in flat background).
Recent measurement of cosmological parameters shows a tendency of favoring slightly positive curvature \cite{Planck-2018-VI, Handley-2019, Park-Ratra-2019, DiValentino-etal-2020, Efstathiou-Gratton-2020, Ellis-Larena-2020}.

Although he missed the opportunity to predict the dynamic universe, Einstein's legacy of establishing modern cosmology over one hundred years ago by introducing the Cosmological Principle and the enigmatic cosmological constant may yet be extended  by his choice of the spherical geometry with closed topology \cite{Einstein-1917}.

%
%

\section*{Acknowledgments}

We thank Professors Gary Gibbons and Anthony Challinor for useful discussions.
J.D.B. was supported by the Science and Technology Facilities Council (STFC)
of the U.K.
J.H.\ was supported by Basic Science Research Program through
the National Research Foundation (NRF) of Korea funded by the Ministry of Science, ICT and Future Planning (No.\ 2018R1A6A1A06024970 and NRF-2019R1A2C1003031).
H.N.\ was supported by the National Research
Foundation of Korea funded by the Korean Government (No.\ 2018R1A2B6002466).

\appendix


\section{Exact solutions for a zero-pressure fluid}

\label{App:MDE-solutions}

Here, we present a complete set of exact solutions including the cases of $%
k^{2}=0$ and $3K$. We consider a zero-pressure fluid ($p=0=\delta p$, $\Pi =0
$) with the cosmological constant and the background curvature. Relatively
decaying solutions are absorbed in the lower bound of integration, and $g(%
\mathbf{x})$ is the remnant gauge mode in the SG.

\begin{tabbing}
\hskip 1.2cm
  \=  General $k^2$ \hskip 5.4cm
  \=  $k^2 = 0$ \hskip 4cm
  \=  $k^2 = 3K$
  \\
---------------------------------------------------------------------------------------------------------------------------------------------------------
  \\
$ v_\chi $
  \>  $ C {c^2 \over a H} \left( 1 + a^2 H \dot H
        \int^t {dt \over \dot a^2} \right) $
  \>  $ C {c^2 \over a H} \left( 1 + a^2 H \dot H
        \int^t {dt \over \dot a^2} \right) $
  \>  $ C {c^2 \over a H} \left( 1 + a^2 H \dot H
        \int^t {dt \over \dot a^2} \right) $
  \\
$ v_\kappa $
  \>  $ C {(k^2 - 3K) c^2 \over
        k^2 c^2 - 3 a^2 \dot H } {c^2 \over a H}
        \left( 1 + a^2 H \dot H \int^t {dt \over \dot a^2}
        \right) $
  \>  $ C {K c^2 \over a^2 \dot H } {c^2 \over a H}
        \left( 1 + a^2 H \dot H \int^t {dt \over \dot a^2}
        \right) $
  \>  $ 0 $
  \\
$ v_\alpha $
  \>  $ g {c \over a} $
  \>  $ g {c \over a} $
  \>  $ g {c \over a} $
  \\
$ v_\varphi $
  \>  $ C {c^2 \over a H} \left( 1 + K c^2 H
        \int^t {dt \over \dot a^2} \right) $
  \>  $ C {c^2 \over a H} \left( 1 + K c^2 H
        \int^t {dt \over \dot a^2} \right) $
  \>  $ C {c^2 \over a H} \left( 1 + K c^2 H
        \int^t {dt \over \dot a^2} \right) $
  \\
$ v_\delta $
  \>  $ - C (k^2 - 3 K) {c^4 \over 3 a} \int^t
        {dt \over \dot a^2} $
  \>  $ C {K c^4 \over a} \int^t
        {dt \over \dot a^2} $
  \>  $ 0 $
  \\
---------------------------------------------------------------------------------------------------------------------------------------------------------
  \\
$ \chi_v $
  \>  $ C {c \over H} \left( 1 + a^2 H \dot H \int^t {dt \over
        \dot a^2} \right) $
  \>  $ C {c \over H} \left( 1 + a^2 H \dot H \int^t {dt \over
        \dot a^2} \right) $
  \>  $ C {c \over H} \left( 1 + a^2 H \dot H \int^t {dt \over
        \dot a^2} \right) $
  \\
$ \chi_\kappa $
  \>  $ C { 12 \pi G \varrho a^2 \over k^2 c^2 - 3 a^2 \dot H }
        {c \over H} \left( 1 + a^2 H \dot H \int^t {dt \over \dot a^2} \right) $
  \>  $ - C { 4 \pi G \varrho \over \dot H }
        {c \over H} \left( 1 + a^2 H \dot H \int^t {dt \over \dot a^2} \right) $
  \>  $ C {c \over H}
        \left( 1 + a^2 H \dot H \int^t {dt \over \dot a^2}
        \right) $
  \\
$ \chi_\alpha $
  \>  $ \chi_v - g $
  \>  $ \chi_v - g $
  \>  $ \chi_v - g $
  \\
$ \chi_\varphi $
  \>  $ - C 4 \pi G \varrho a^2 c \int^t {dt \over \dot a^2} $
  \>  $ - C 4 \pi G \varrho a^2 c \int^t {dt \over \dot a^2} $
  \>  $ - C 4 \pi G \varrho a^2 c \int^t {dt \over \dot a^2} $
  \\
$ \chi_\delta $
  \>  $ C {c \over H} \left[ 1 + \left( {k^2 c^2 \over 3}
        - 4 \pi G \varrho a^2 \right) H \int^t {dt \over \dot a^2}
        \right] $
  \>  $ C {c \over H} \left( 1
        - 4 \pi G \varrho a^2 H \int^t {dt \over \dot a^2}
        \right) $
  \>  $ C {c \over H}
        \left( 1 + a^2 H \dot H \int^t {dt \over \dot a^2}
        \right) $
  \\
---------------------------------------------------------------------------------------------------------------------------------------------------------
  \\
$ \kappa_v $
  \>  $ C {(k^2 - 3K) c^2 \over a^2 H}
        \left( 1 + a^2 H \dot H \int^t
        {dt \over \dot a^2} \right) $
  \>  $ - C {3 K c^2 \over a^2 H}
        \left( 1 + a^2 H \dot H \int^t
        {dt \over \dot a^2} \right) $
  \>  $ 0 $
  \\
$ \kappa_\chi $
  \>  $ - C {12 \pi G \varrho \over H}
        \left( 1 + a^2 H \dot H \int^t {dt \over \dot a^2}
        \right) $
  \>  $ - C {12 \pi G \varrho \over H}
        \left( 1 + a^2 H \dot H \int^t {dt \over \dot a^2}
        \right) $
  \>  $ - C {12 \pi G \varrho \over H}
        \left( 1 + a^2 H \dot H \int^t {dt \over \dot a^2}
        \right) $
  \\
$ \kappa_\alpha $
  \>  $ \kappa_v + g {1 \over c} \left( 3 \dot H
        - {k^2 c^2 \over a^2} \right) $
  \>  $ \kappa_v + g {1 \over c} 3 \dot H $
  \>  $ - g {1 \over c} 12 \pi G \varrho $
  \\
$ \kappa_\varphi $
  \>  $ - C {4 \pi G \varrho \over H}
        \left( 3 + k^2 c^2 H \int^t {dt \over \dot a^2} \right) $
  \>  $ - C {12 \pi G \varrho \over H} $
  \>  $ - C {12 \pi G \varrho \over H}
        \left( 1 + K c^2 H \int^t {dt \over \dot a^2} \right) $
  \\
$ \kappa_\delta $
  \>  $ C {(k^2 - 3K) c^2 \over a^2 H}
        \left( 1 + {1 \over 3} k^2 c^2 H \int^t
        {dt \over \dot a^2} \right) $
  \>  $ - C {3 K c^2 \over a^2 H} $
  \>  $ 0 $
  \\
---------------------------------------------------------------------------------------------------------------------------------------------------------
  \\
$ \alpha_v $
  \>  $ 0 $
  \>  $ 0 $
  \>  $ 0 $
  \\
$ \alpha_\chi $
  \>  $ - C 4 \pi G \varrho a^2 H \int^t {dt \over \dot a^2} $
  \>  $ - C 4 \pi G \varrho a^2 H \int^t {dt \over \dot a^2} $
  \>  $ - C 4 \pi G \varrho a^2 H \int^t {dt \over \dot a^2} $
  \\
$ \alpha_\kappa $
  \>  $ - C {(k^2 - 3 K) c^2 \over
        (k^2 c^2 - 3 a^2 \dot H)^2} 4 \pi G \varrho a^2
        \left( 3 + k^2 c^2 H \int^t {dt \over \dot a^2} \right) $
  \>  $ C {K c^2 \over a^2} {4 \pi G \varrho \over \dot H} $
  \>  $ 0 $
  \\
$ \alpha_\varphi $
  \>  $ - C {4 \pi G \varrho \over H^2} $
  \>  $ - C {4 \pi G \varrho \over H^2} $
  \>  $ - C {4 \pi G \varrho \over H^2} $
  \\
$ \alpha_\delta $
  \>  $ C {(k^2 - 3K) c^2 \over 3 a^2 H^2} $
  \>  $ - C {K c^2 \over a^2 H^2} $
  \>  $ 0 $
  \\
---------------------------------------------------------------------------------------------------------------------------------------------------------
  \\
$ \varphi_v $
  \>  $ C \left( 1 + K c^2 H \int^t {dt \over \dot a^2} \right) $
  \>  $ C \left( 1 + K c^2 H \int^t {dt \over \dot a^2} \right) $
  \>  $ C \left( 1 + K c^2 H \int^t {dt \over \dot a^2} \right) $
  \\
$ \varphi_\chi $
  \>  $ C 4 \pi G \varrho a^2 H \int^t {dt \over \dot a^2} $
  \>  $ C 4 \pi G \varrho a^2 H \int^t {dt \over \dot a^2} $
  \>  $ C 4 \pi G \varrho a^2 H \int^t {dt \over \dot a^2} $
  \\
$ \varphi_\kappa $
  \>  $ C { 4 \pi G \varrho a^2 \over
        k^2 c^2 - 3 a^2 \dot H } \left( 3 + k^2 c^2 H
        \int^t {dt \over \dot a^2} \right) $
  \>  $ - C { 4 \pi G \varrho \over \dot H } $
  \>  $ C \left( 1 + K c^2 H
        \int^t {dt \over \dot a^2} \right) $
  \\
$ \varphi_\alpha $
  \>  $ \varphi_v - g {H \over c} $
  \>  $ \varphi_v - g {H \over c} $
  \>  $ \varphi_v - g {H \over c} $
  \\
$ \varphi_\delta $
  \>  $ C \left( 1 + {1 \over 3} k^2 c^2 H
        \int^t {dt \over \dot a^2} \right) $
  \>  $ C $
  \>  $ C \left( 1 + K c^2 H
        \int^t {dt \over \dot a^2} \right) $
  \\
---------------------------------------------------------------------------------------------------------------------------------------------------------
  \\
$ \delta_v $
  \>  $ C (k^2 - 3K) c^2 H \int^t {dt \over \dot a^2} $
  \>  $ - 3 C K c^2 H \int^t {dt \over \dot a^2} $
  \>  $ 0 $
  \\
$ \delta_\chi $
  \>  $ C \left[ 3 + \left( k^2 c^2
        - 12 \pi G \varrho a^2 \right) H
        \int^t {dt \over \dot a^2} \right] $
  \>  $ 3 C \left( 1 - 4 \pi G \varrho a^2 H
        \int^t {dt \over \dot a^2} \right) $
  \>  $ 3 C \left( 1 + a^2 H \dot H
        \int^t {dt \over \dot a^2} \right) $
  \\
$ \delta_\kappa $
  \>  $ C {(k^2 - 3K) c^2 \over k^2 c^2 - 3 a^2 \dot H}
        \left( 3 + k^2 c^2 H \int^t
        {dt \over \dot a^2} \right) $
  \>  $ C {3K c^2 \over a^2 \dot H} $
  \>  $ 0 $
  \\
$ \delta_\alpha $
  \>  $ \delta_v + 3 g {H \over c} $
  \>  $ \delta_v + 3 g {H \over c} $
  \>  $ 3 g {H \over c} $
  \\
$ \delta_\varphi $
  \>  $ C \left( 3 + k^2 c^2 H \int^t {dt \over \dot a^2}
        \right) $
  \>  $ 3 C $
  \>  $ 3 C \left( 1 + K c^2 H \int^t {dt \over \dot a^2}
        \right) $
  \\
---------------------------------------------------------------------------------------------------------------------------------------------------------
\end{tabbing}


\section{Taub constraint}

\label{App:Taub-constraint}


\subsection{Derivation}

Einstein's equations are
\begin{equation}
\widetilde{E}^{ab}\equiv \widetilde{R}^{ab}-{\frac{1}{2}}\widetilde{g}^{ab}%
\widetilde{R}+\Lambda \widetilde{g}^{ab}-{\frac{8\pi G}{c^{4}}}\widetilde{T}%
^{ab}=0.  \label{E^ab}
\end{equation}%
To second order in perturbation, the metric tensor and its inverse are
\begin{equation}
\widetilde{g}_{ab}\equiv g_{ab}+h_{ab},\quad \widetilde{g}%
^{ab}=g^{ab}-h^{ab}+h_{c}^{a}h^{cb},
\end{equation}%
where $h_{ab}$ includes the second order and its indices are raised and
lowered using the background metric $g_{ab}$ and its inverse metric $g^{ab}$%
. The connection is
\begin{equation}
\widetilde{\Gamma }_{bc}^{a}=\Gamma _{bc}^{a}+{\frac{1}{2}}\left(
h_{b:c}^{a}+h_{c:b}^{a}-h_{bc}^{\;\;\;:a}\right) -{\frac{1}{2}}h^{ad}\left(
h_{bd:c}+h_{cd:b}-h_{bc:d}\right) ,
\end{equation}%
where a colon indicates the covariant derivative using the background metric
$g_{ab}$. The curvatures are
\begin{eqnarray}
&&\widetilde{R}_{\;\;bcd}^{a}=R_{\;\;bcd}^{a}+h_{b:[dc]}^{a}+h_{[d:\{b%
\}c]}^{a}-h_{b[d\;\;\;c]}^{\;\;\;\;\;:a}-h^{ae}\left(
h_{eb:[dc]}+h_{e[d:\{b\}c]}-h_{b[d:\{e\}c]}\right)   \notag \\
&&\qquad +{\frac{1}{2}}\left(
h_{b:[d}^{e}+h_{[d:\{b\}}^{e}-h_{b[d}^{\;\;\;\;\;:e}\right) \left(
h_{c]:e}^{a}-h_{\{e\}:c]}^{a}-h_{c]e}^{\;\;\;\;:a}\right) ,  \notag \\
&&\widetilde{R}_{ab}=R_{ab}+{\frac{1}{2}}\left(
h_{a:bc}^{c}+h_{b:ac}^{c}-h_{ab\;\;\;c}^{\;\;\;\;:c}-h_{:ab}\right) -{\frac{1%
}{2}}h^{ce}\left( h_{ea:bc}+h_{eb:ac}-h_{ec:ab}-h_{ab:ec}\right)   \notag \\
&&\qquad +{\frac{1}{4}}\left(
h_{a:b}^{e}+h_{b:a}^{e}-h_{ab}^{\;\;\;:e}\right) \left(
h_{,e}-2h_{e:c}^{c}\right) +{\frac{1}{4}}h_{\;\;\;:a}^{cd}h_{cd:b}+{\frac{1}{%
2}}h_{a}^{c:d}\left( h_{bc:d}-h_{bd:c}\right) \equiv
R_{ab}+R_{ab}^{L}+R_{ab}^{Q},  \notag \\
&&\widetilde{R}=R-h^{ab}R_{ab}+h_{\;\;\;\;:ab}^{ab}-h_{\;\;%
\;a}^{:a}+h_{c}^{a}h^{cb}R_{ab}+h^{ab}\left(
-h_{a:cb}^{c}+h_{ab\;\;\;c}^{\;\;\;\;:c}-h_{a:bc}^{c}+h_{,a:b}\right)
\notag \\
&&\qquad -h_{\;\;\;\;:b}^{ab}h_{a:c}^{c}+h_{\;\;\;\;:b}^{ab}h_{,a}-{\frac{1}{%
4}}h^{:a}h_{,a}+{\frac{1}{4}}h^{ab;c}\left( 3h_{ab:c}-2h_{ac:b}\right)
\equiv R+R^{L}+R^{Q},
\end{eqnarray}%
where $h\equiv h_{c}^{c}$ and we have $A_{[ab]}\equiv $  $\frac{1}{2}%
(A_{ab}-A_{ba})$. The indices $L$ and $Q$ indicate the linear and quadratic
parts, respectively. The quadratic part is the terms with quadratic
combination of two first (linear) order terms. The linear part can be
decomposed into the first-order and second-order perturbations, like $%
R_{ab}^{L}=R_{ab}^{(1)}+R_{ab}^{(2)}$; for example, to the second order, we
have $h_{ab}=h_{ab}^{(1)}+h_{ab}^{(2)}\equiv h_{ab}^{L}$.

The background and first order Einstein equation give $E^{ab}=0$ and $%
E^{(1)ab}=0$. The equation to the second order can be arranged as
\begin{equation}
E^{Lab}\equiv \widetilde{R}^{Lab}-{\frac{1}{2}}g^{ab}\widetilde{R}%
^{L}+h^{ab}\left( {\frac{1}{2}}R-\Lambda \right) -{\frac{8\pi G}{c^{4}}}%
T^{Lab}=-E^{Qab}\equiv {\frac{8\pi G}{c^{4}}}t^{ab},  \label{E^Qab}
\end{equation}%
with
\begin{eqnarray}
&&{\frac{8\pi G}{c^{4}}}t^{ab}={\frac{1}{2}}h^{cd}\left(
h_{c\;\;\;d}^{a:b}+h_{c\;\;\;d}^{b:a}-h_{cd}^{\;\;\;\;:ab}-h_{\;\;\;%
\;:cd}^{ab}\right) -{\frac{1}{4}}\left( h^{ca:b}+h^{cb:a}-h^{ab:c}\right)
\left( h_{,c}-2h_{c:d}^{d}\right) -{\frac{1}{4}}h^{cd:a}h_{cd}^{\;\;\;\;:b}
\notag \\
&&\qquad -{\frac{1}{2}}h^{ac:d}\left( h_{c:d}^{b}-h_{d:c}^{b}\right)
+h^{c(a}\left(
h_{c\;\;\;\;d}^{d:b)}+h_{\;\;\;\;\;:cd}^{b)d}-h_{c\;\;\;\;d}^{b):d}-h_{\;\;%
\;c}^{:b)}\right) -h^{ac}h^{bd}R_{cd}-2h^{cd}h_{c}^{(a}R_{d}^{b)}  \notag \\
&&\qquad +{\frac{1}{2}}g^{ab}\Big[h_{e}^{c}h^{ed}R_{cd}-h^{cd}\left(
h_{c:de}^{e}+h_{c:ed}^{e}-h_{cd\;\;\;e}^{\;\;\;\;:e}-h_{,c:d}\right) -{\frac{%
1}{4}}h^{:c}h_{,c}+h^{:c}h_{c:d}^{d}-h_{\;\;\;\;:d}^{cd}h_{c:e}^{e}  \notag
\\
&&\qquad +{\frac{1}{4}}h^{cd:e}\left( 3h_{cd:e}-2h_{de:c}\right) \Big]+{%
\frac{1}{2}}h^{ab}\left(
h^{cd}R_{cd}+h_{\;\;\;c}^{:c}-h_{\;\;\;\;:cd}^{cd}\right)
+h_{c}^{a}h^{bc}\left( {\frac{1}{2}}R-\Lambda \right) +{\frac{8\pi G}{c^{4}}}%
T^{Qab}.  \label{t_ab}
\end{eqnarray}%
This was presented by Taub in Eq.\ (3.5) of \cite{Taub-1961} for the
Minkowski background. Here we consider a general background metric $g_{ab}$.

From $\widetilde{E}_{\;\;\;\;;b}^{ab}\equiv 0$, we have $E_{\;\;\;%
\;:b}^{ab}=0=E_{\;\;\;\;\;\;\;\;:b}^{(1)ab}$ and $E_{\;\;\;\;\;\;\;%
\;:b}^{(2)ab}=0$. Thus $E_{\;\;\;\;\;\;:b}^{Lab}=0$, and we have $%
E_{\;\;\;\;\;\;:b}^{Qab}=0=t_{\;\;\;:b}^{ab}$. For a Killing vector $\xi _{a}
$, where $\xi _{a:b}+\xi _{b:a}\equiv 0$, we have
\begin{equation}
0=\left( \sqrt{-g}t^{ab}\xi _{b}\right) _{:a}=\left( \sqrt{-g}t^{ab}\xi
_{b}\right) _{,a}.
\end{equation}%
thus (see Eq.\ (4.7) of \cite{Taub-1961})
\bea
   & & 0 = \int \left( \sqrt{-g} t^{ab} \xi_b \right)_{:a} d^4 x
       = \int \sqrt{-g} t^{ab} \xi_b d^3 \sigma_a
       = \int \sqrt{-g} t^{ab} \xi_b n_a d^3 x,
\eea
with $n_a$ the timelike normal ($n_i \equiv 0$) four vector.
Therefore, we define
\begin{equation}
\mathcal{T}\equiv -{\frac{8\pi G}{c^{4}}}\int \sqrt{-g}t^{0b}\xi
_{b}d^{3}x=\int \sqrt{-g}E^{Q0b}\xi _{b}d^{3}x=0,  \label{Taub}
\end{equation}
and call this the Taub constraint. In the presence of the Killing vectors
in the background metric $g_{ab}$, Fischer, Marsden and Moncrief \cite{FMM,
mon}, have proved the violation of this condition as the criterion of
linearization instability for the \textit{vacuum} case. Similar results hold
for Einstein field equations coupled with matter fields such as scalar
fields, electromagnetic fields and Yang-Mills fields \cite{arms-marsden,
arms, arms2}.


\subsection{ADM constraint formulation}

Evaluation of Eq.\ (\ref{Taub}) with Eq.\ (\ref{t_ab}), needs complicated
algebra. There is a simpler formulation using the constraint equations.
The ADM (Arnowitt-Deser-Misner) energy and momentum constraint equations can
be written as (Eq.\ (3.14) in \cite{Arnowitt-Deser-Misner-1962}),
\begin{eqnarray}
&&\mathcal{E}^{0}\equiv K^{ij}K_{ij}-K^{2}-R^{(h)}+{\frac{16\pi G}{c^{4}}}%
E+2\Lambda =0,  \label{E-constraint-H} \\
&&\mathcal{E}^{i}\equiv K_{\;\;\;\parallel j}^{ij}-K^{\parallel i}-{\frac{%
8\pi G}{c^{4}}}J^{i}=0.  \label{Mom-constraint-H}
\end{eqnarray}%
The indices and the covariant derivatives ($\parallel $) in the ADM notation
are based on the ADM metric $h_{ij}\equiv \widetilde{g}_{ij}$. From Eq.\ (%
\ref{E^ab}), we can show
\begin{equation}
\widetilde{E}^{00}=-{\frac{1}{2N^{2}}}\mathcal{E}^{0},\quad \widetilde{E}%
^{0i}={\frac{1}{N}}\mathcal{E}^{i}+{\frac{N^{i}}{4N^{4}}}\mathcal{E}^{0}.
\end{equation}%
To the second order, we have
\begin{equation}
\widetilde{E}^{00}=E^{(0)00}+E^{L00}+E^{Q00},\quad \widetilde{E}%
^{0i}=E^{L0i}+E^{Q0i};\quad \mathcal{E}^{0}\equiv \mathcal{E}^{(0)0}+%
\mathcal{E}^{L0}+\mathcal{E}^{Q0},\quad \mathcal{E}^{i}\equiv \mathcal{E}%
^{Li}+\mathcal{E}^{Qi}.
\end{equation}%
As we have $E^{(0)00}=0=\mathcal{E}^{(0)0}$ for the background, and $%
E^{(1)00}=0=\mathcal{E}^{(1)0}$ and $E^{(1)0i}=0=\mathcal{E}^{(1)i}$ for the
first-order perturbation, the quadratic parts become
\begin{equation}
E^{Q00}=-{\frac{1}{2(N^{(0)})^{2}}}\mathcal{E}^{Q0},\quad E^{Q0i}={\frac{1}{%
N^{(0)}}}\mathcal{E}^{Qi}.  \label{EQ-relations}
\end{equation}%
Using this Eq.\ (\ref{Taub}) gives for the Taub constraint,
\begin{equation}
\mathcal{T}=\int \sqrt{-g}\xi _{b}E^{Q0b}d^{3}x=\int \sqrt{h^{(0)}}%
N^{(0)}\left( \xi _{0}E^{Q00}+\xi _{i}E^{Q0i}\right) d^{3}x=\int \sqrt{%
h^{(0)}}\left( -{\frac{1}{2N^{(0)}}}\xi _{0}\mathcal{E}^{Q0}+\xi _{i}%
\mathcal{E}^{Qi}\right) d^{3}x,  \label{Taub-2}
\end{equation}%
where we used $\sqrt{-g}=N^{(0)}\sqrt{h^{(0)}}$. This is an alternative
presentation of the Taub constraint to Eq.\ (\ref{Taub}) which needs only
the energy and momentum constraint equations.

In the cosmological background, the Taub constraint derived in Eq.\ (\ref%
{Taub-2}) yields
\begin{equation}
\mathcal{T}=\int \sqrt{\gamma }%
a^{4}\left( \xi _{0}E^{Q00}+\xi _{i}E^{Q0i}\right) d^{3}x=\int \sqrt{\gamma }%
a^{3}\left( -{\frac{1}{2a}}\xi _{0}\mathcal{E}^{Q0}+\xi _{i}\mathcal{E}%
^{Qi}\right) d^{3}x,  \label{Taub-cosmology-2}
\end{equation}%
where $\gamma $ is the determinant of $\gamma _{ij}$. The Friedmann metric
has six space-like Killing vectors \cite{Ray-Zimmerman-1977}.
Einstein's static model has an additional timelike Killing vector with $\xi
^{a}\equiv \delta _{0}^{a}$. We will consider the Taub constraint based
on this timelike Killing vector. Using $\xi _{a}=-\delta _{a}^{0}$,
Eq.\ (\ref{Taub-cosmology-2}) gives
\begin{equation}
\mathcal{T}=\int \sqrt{-g}\xi _{b}E^{Q0b}d^{3}x=-a_{0}^{4}\int
\sqrt{\gamma }E^{Q00}d^{3}x={\frac{1}{2}}a_{0}^{2}\int \sqrt{\gamma }%
\mathcal{E}^{Q0}d^{3}x.  \label{Taub-cosmology-3}
\end{equation}%
Thus, for evaluation of the Taub constraint in our case, we only
need the energy constraint equation to second order.


\subsection{The energy constraint equation to second order}

The fully nonlinear and exact perturbation equations in the presence of
background curvature were presented in \cite{Noh-2014}; the equations are
derived by taking a spatial gauge $\gamma \equiv 0$ in the metric in Eq.\ (%
\ref{metric}), and replacing $a\beta _{,i}$ and $-v_{,i}$ by $\chi _{i}$ and
$v_{i}$, respectively, now including the vector-type perturbation as well.
The ADM energy constraint equation gives (Eq.\ (3.2) in \cite{Noh-2014}):
\begin{eqnarray}
&&\mathcal{E}^{0}=-{\frac{6}{c^{2}}}\left( H^{2}-{\frac{8\pi G}{3}}%
\widetilde{\varrho }+{\frac{Kc^{2}}{a^{2}(1+2\varphi )}}-{\frac{\Lambda c^{2}%
}{3}}\right) +{\frac{4}{c^{2}}}H\kappa +{\frac{4\Delta \varphi }{%
a^{2}(1+2\varphi )^{2}}}  \notag \\
&&\qquad -{\frac{2}{3c^{2}}}\kappa ^{2}+{\frac{16\pi G}{c^{2}}}\left(
\widetilde{\varrho }+{\frac{\widetilde{p}}{c^{2}}}\right) (\gamma ^{2}-1)-{%
\frac{6\varphi ^{|i}\varphi _{,i}}{a^{2}(1+2\varphi )^{3}}}+\overline{K}%
_{j}^{i}\overline{K}_{i}^{j},  \label{eq2-FNL}
\end{eqnarray}%
with $\gamma $ the Lorentz factor, and $N$ the lapse function, where
\begin{eqnarray}
&&\gamma \equiv {\frac{1}{\sqrt{1-{\frac{v^{k}v_{k}}{c^{2}(1+2\varphi )}}}}}%
,\quad N\equiv a\mathcal{N}\equiv a\sqrt{1+2\alpha +{\frac{\chi ^{k}\chi _{k}%
}{a^{2}(1+2\varphi )}}},\quad \overline{K}_{j}^{i}\overline{K}_{i}^{j}={%
\frac{1}{a^{4}\mathcal{N}^{2}(1+2\varphi )^{2}}}\Bigg\{{\frac{1}{2}}\chi
^{i|j}\left( \chi _{i|j}+\chi _{j|i}\right)  \notag \\
&&-{\frac{1}{3}}\chi _{\;\;|i}^{i}\chi _{\;\;|j}^{j}-{\frac{4}{1+2\varphi }}%
\left[ {\frac{1}{2}}\chi ^{i}\varphi ^{|j}\left( \chi _{i|j}+\chi
_{j|i}\right) -{\frac{1}{3}}\chi _{\;\;|i}^{i}\chi ^{j}\varphi _{,j}\right] +%
{\frac{2}{(1+2\varphi )^{2}}}\left( \chi ^{i}\chi _{i}\varphi ^{|j}\varphi
_{,j}+{\frac{1}{3}}\chi ^{i}\chi ^{j}\varphi _{,i}\varphi _{,j}\right) %
\Bigg\}.  \label{K-bar-eq}
\end{eqnarray}%
To second order, we have
\begin{eqnarray}
&&\mathcal{E}^{0}=-{\frac{6}{c^{2}}}\left( H^{2}-{\frac{8\pi G}{3}}\varrho +{%
\frac{Kc^{2}}{a^{2}}}-{\frac{\Lambda c^{2}}{3}}\right) +{\frac{4}{c^{2}}}%
\left( 4\pi G\delta \varrho +H\kappa +c^{2}{\frac{\Delta +3K}{a^{2}}}\varphi
\right)  \notag \\
&&\qquad +{\frac{16\pi G}{c^{2}}}\left( \varrho +{\frac{p}{c^{2}}}\right) {%
\frac{v^{i}v_{i}}{c^{2}}}-{\frac{2}{3c^{2}}}\kappa ^{2}-{\frac{2}{a^{2}}}%
\left[ 3\varphi ^{|i}\varphi _{,i}+4\varphi \left( 2\Delta +3K\right)
\varphi \right] +{\frac{1}{a^{4}}}\left[ {\frac{1}{2}}\chi ^{i|j}\left( \chi
_{i|j}+\chi _{j|i}\right) -{\frac{1}{3}}\chi _{\;\;|i}^{i}\chi _{\;\;|j}^{j}%
\right]  \notag \\
&&\qquad \equiv \mathcal{E}^{(0)0}+\mathcal{E}^{L0}+\mathcal{E}^{Q0}.
\label{eq2-E0-scalar}
\end{eqnarray}%
For the scalar perturbation, we have $v_{i}\equiv -v_{,i}$ and $\chi
_{i}\equiv \chi _{,i}$. The evaluation of the Taub constraint in Eq.\ (\ref%
{Taub-cosmology-3}) using Eq.\ (\ref{eq2-E0-scalar}) in a couple of gauge
conditions in Einstein's static model is presented in Section \ref%
{Taub-constraint}.

\end{widetext}
%
%


\end{document}